\documentstyle[psfig,amssym,aps,preprint]{revtex}

\newcommand{\bea}{\begin{eqnarray}}
\newcommand{\eea}{\end{eqnarray}}
\newcommand{\be}{\begin{equation}}
\newcommand{\ee}{\end{equation}}
\newcommand{\bm}[1]{\mbox{\boldmath $#1$}}
\newcommand{\nn}{\nonumber}
\newcommand{\ds}{\displaystyle}
\newcommand{\Tr}{{\rm Tr}}

\newcommand{\B}{\mbox{\tiny $B$}}
\newcommand{\sample}[1]{\!\!\!\!\raisebox{-0.8ex}{
                        \scriptsize $\mathcal{#1}$}}
\newcommand{\sampleb}[1]{\!\!\!\!\raisebox{-0.8ex}{
                         \scriptsize $\bar{\mathcal{#1}}$}}
\newcommand{\samplen}[1]{\!\!\!\!\raisebox{-0.8ex}{
                         \scriptsize $#1$}}
\newcommand{\<}{\langle\,}
\renewcommand{\>}{\,\rangle}

\draft
\makeatletter

                \def\@preprint{}
                \def\preprint#1#2#3{%
                \ifpreprintsty
                \def\@preprint{
                \noindent \hbox{#1}\hfill\hbox{#2}\\
                \hbox{#3}\vskip -2ex}%
                \fi
                }

\begin{document}

%
%

\preprint{hep-ph/9701347}{PITHA 97/4}{}
\title{
       QCD Corrections to Decay Distributions
       of Neutral Higgs Bosons with (In)definite CP Parity
        }
\author{
        W. Bernreuther,
        A. Brandenburg\footnote{supported  by Deutsche
                                Forschungsgemeinschaft.},
        and M. Flesch\footnote{supported  by BMBF
                               contract 057AC9EP.}
        }
\address{
        Institut f\"ur Theoretische Physik, RWTH Aachen, 
        52056 Aachen, Germany
        }
\date{\today}
\maketitle{}
\begin{abstract}
We compute the order $\alpha_s$ QCD corrections to the density 
matrix for the decay of a neutral Higgs boson $\varphi$ with (in)definite 
CP parity into a quark antiquark pair, respectively the QED corrections 
for the decay into a pair of charged leptons. We classify and calculate 
single spin asymmetries and spin-spin correlations which are generated 
by the scalar and pseudoscalar Yukawa couplings. These spin effects can 
be traced in $\varphi \to \tau^-\tau^+$ and, for heavy Higgs bosons, 
in  $\varphi \to t\bar{t}$. We also calculate resulting correlations 
among the final states and estimate, for the respective decay modes, 
the number of events needed to measure the Yukawa couplings with these 
correlations at the 3$\sigma$ level.
\end{abstract}

\newpage

%
%

\section{Introduction}
Many extensions of the Standard Model (SM) involve more than one scalar 
field multiplet and thus  predict the existence of more than just one 
Higgs boson. In particular, quite a number of these models -- including 
the 2 Higgs doublet extensions of the SM \cite{Lee,Bran,Wolf,Wein} -- 
allow for the violation of CP symmetry  by the scalar self interactions.
This type of CP violation is of great interest for scenarios that attempt 
to explain the baryon asymmetry of the universe \cite{Cohen}.\par
A manifestation  of CP violation in the scalar sector would be the
existence of neutral Higgs bosons $\varphi$  of undefined CP parity, 
i.e., Higgs bosons having both scalar and pseudoscalar couplings to quarks 
and leptons. If neutral  Higgs boson(s) should be discovered in the 
future, one would eventually like to know its (their) P and CP quantum 
numbers. In this context several proposals and theoretical studies
have been made in the literature. For instance if $\varphi$ has both 
scalar and pseudoscalar Yukawa couplings then a CP-violating spin-spin 
correlation is induced already at tree level in the decays of $\varphi$ 
to fermion-antifermion pairs. This spin-spin correlation could be traced in
 $\varphi\to\tau^-\tau^+$ and $\varphi\to t\bar{t}$ \cite{BBra}.
(For related proposals, see \cite{Nelson,Nor,Chang,HMK,Kremer,Seg1,Grad}.)
The modes $\varphi \to W^+W^-, ZZ$ could also be employed to infer the
parity and CP properties \cite{comm} of $\varphi$ 
\cite{Chang,Kremer,Xu,DK,Seg2,CoWi}.
Other reactions that may be used for this purpose are associated 
 $t {\bar t}\varphi $ production \cite{Soni,GGHe}, $\varphi$ production by 
high energetic photon photon collisions \cite{Kremer,GG,ABB}, and by
high energetic $\mu^-\mu^+$ annihilation \cite{ASoni,Pil}.\par
In this article we calculate the order $\alpha_s$ QCD corrections
to the density matrix for the decay of a neutral Higgs boson $\varphi$
with arbitrary scalar and pseudoscalar Yukawa couplings 
 into a quark antiquark pair. (QCD corrections 
to the decay width of a scalar and/or pseudoscalar $\varphi\to q{\bar q}$ 
were computed in \cite{Braaten,Saka,Larin,Chet,Djou,Kniehl}.)
We classify and calculate single spin asymmetries and spin-spin
correlations which are generated by the scalar and/or pseudoscalar 
Yukawa  couplings. These spin effects can be traced for heavy Higgs bosons
in  $\varphi \to  t\bar{t}$ because top quarks decay before they hadronize 
and because they auto-analyze their spins through their parity-violating 
weak decays. We show that there are two correlations among the final 
states from $ t\bar{t}$ decay whose combined use allows to investigate 
the CP property of a neutral $\varphi$.  \par
We then apply these results to the decay $\varphi\to\tau^-\tau^+$
and its QED corrections. We analyze correlations for $\tau$ decay modes 
that have the best $\tau$ spin analyzer quality.  Finally we
estimate, for the respective $\varphi$ decay modes, the number of
events needed to measure the top and $\tau$ Yukawa couplings with these 
correlations at the 3$\sigma$ level.

\section{The decay \lowercase{$\varphi\to f\bar{f}$} $\!X$}

Let us briefly recapitulate the salient features of the simplest models
that predict neutral Higgs bosons with undefined CP parity; 
these are  2 Higgs doublet extensions of the SM with natural flavor 
conservation at the tree level (see, e.g., \cite{Wein}). Explicit CP 
violation in the Higgs potential leads to three physical neutral boson 
states with scalar and pseudoscalar couplings to fermions. In the following
 $\varphi$ denotes one of these bosons. The Yukawa interactions with a 
quark or lepton field $f$ read:
\begin{equation} 
{\cal L}_{Y}=-\frac{m_{f}}{v}\bar{f}(a_{f}+i\gamma_{5}\tilde{a}_{f}) f
\varphi\,\, , 
\label{Ly}
\end{equation}
where $v=(\sqrt{2}G_F)^{-1/2}$ and $G_F$ is Fermi's constant, $m_f$ is the 
fermion mass, $a_{f}$ and $\tilde{a}_{f}$ are scalar and pseudoscalar 
coupling constants, respectively. If $a_{f}\tilde{a}_{f}\neq 0$ then 
${\cal L}_{Y}$ is CP-violating. \par
For the reaction $\varphi\to f\bar{f} X$ the spin density matrix of the
 $f\bar{f}$ subsystem is defined in the $f\bar{f}$ center of mass system by:
\bea
R_{\alpha_1\alpha_2,\ \beta_1\beta_2}({\bf k})=\sum_{{\rm X}}&&  
\<f(k_1,\alpha_1),\bar f(k_2,\beta_1),X|{\cal T}|\varphi(q)\>\nn\\
&&\<f(k_1,\alpha_2),\bar f(k_2,\beta_2),X|{\cal T}|\varphi(q)\>^{*}\,\,.
\label{Rdef}
\eea 
Here ${\bf k}={\bf k}_{1}$ is the $f$ momentum in the $f\bar{f}$ c.m. 
frame and $\alpha$ and $\beta$ are spin indices. The sum in (\ref{Rdef}) 
is taken over all discrete and continuous degrees of freedom of the 
unobserved part $X$ of the final state. The matrix (\ref{Rdef}) can be 
decomposed in the spin spaces of $f$ and $\bar{f}$ as follows:
\be
R = A\, 1\!{\rm l}\otimes 1\!{\rm l}+B^{+}_{i}\, {\bm \sigma}^i
\!\otimes 1\!{\rm l}+B^{-}_{i}\, 1\!{\rm l}\otimes{\bm \sigma}^i
+C_{ij}\, {\bm \sigma}^i\!\otimes{\bm \sigma}^j.
\label{Rzer}
\ee
The first (second) factor in the tensor products of the $2\times 2$
unit matrix $1\!{\rm l}$ and of the Pauli matrices ${\bm \sigma}^i$ refers 
to the $f (\bar{f})$ spin space. Because of rotational invariance, the 
functions $B^{\pm}_{i}$ and $C_{ij}$ can be further decomposed:
\bea
B^{\pm}_{i}&=& b_{\pm}\hat{k}_{i} \,\, ,\nn\\
C_{ij}&=& c_{1}\delta_{ij} + c_{2}\hat{k}_{i} \hat{k}_{j} 
+ c_{3}\epsilon_{ijl}\hat{k}_{l} \,\,,
\eea
with $\bf{\hat k}={\bf k}/|{\bf k}|$ and coefficients $A, b_{\pm}$ and 
 $c_{i}$, which only depend on the masses and couplings. These coefficients
can all be separately measured by using suitable observables built from the
spin operators of $f$ and $\bar{f}$. The trace of $R$, $\Tr\{R\}=4A$ is a 
Lorentz scalar, and is related to the decay rate 
$\Gamma(\varphi\to f\bar{f}X)$ by:
\be
\Gamma=\frac{1}{2m_\varphi}\frac{\beta}{8\pi}\Tr\{R\}\,\, ,
\label{Gb}
\ee
where $\beta=(1-4m_{f}^{2}/m_\varphi^2)^{1/2}$ is the velocity of the 
fermion in the $\varphi$ rest frame. One may construct a complete set of 
observables that determine the other coefficients:
\bea 
{\cal O}_{0} &=& {\bf \hat k}\cdot({\bf s}_{1} + {\bf s}_{2}) \,\,,\nn\\
{\cal O}_{1} &=& {\bf \hat k}\cdot({\bf s}_{1} - {\bf s}_{2}) \,\,,\nn\\
{\cal O}_{2} &=& {\bf \hat k}\cdot({\bf s}_{1}\times{\bf s}_{2}) \,\,,\nn\\
{\cal O}_{3} &=& {\bf s}_{1}\cdot{\bf s}_{2} \,\,,\nn\\ 
{\cal O}_{4} &=& ({\bf \hat k}\cdot{\bf s}_{1})\,
({\bf \hat k}\cdot{\bf s}_{2}) \,\, ,
\label{obs}
\eea
where ${\bf s}_{1}=\frac{1}{2}{\bm \sigma}\otimes1\!{\rm l}$ and
 ${\bf s}_{2}=\frac{1}{2}1\!{\rm l}\otimes{\bm \sigma}$ are the spin
operators of $f$ and $\bar{f}$. The expectation values of the observables 
are given by:
\be
\<{\cal O}_{i}\>=\frac{\Tr\{R\cdot {\cal O}_{i}\}}{\Tr\{R\}}
\,\, . \label{erw1}
\ee
These expectation values are trivially related to the coefficients of the
density matrix (\ref{Rdef}): 
\bea
\<{\cal O}_{0}\>&=&\frac{\ds 2 (b_{+}+b_{-})}{\ds 4A} \,\,,\nn\\
\<{\cal O}_{1}\>&=&\frac{\ds 2 (b_{+}-b_{-})}{\ds 4A} \,\,,\nn\\
\<{\cal O}_{2}\>&=&\frac{\ds 2 c_{3}}{\ds 4A} \,\,,\nn\\
\<{\cal O}_{3}\>&=&\frac{\ds 3 c_{1}+c_{2}}{\ds 4A} \,\,,\nn\\
\<{\cal O}_{4}\>&=&\frac{\ds c_{1}+c_{2}}{\ds 4A}\,\,.
\label{exp}
\eea
In Table 1 we give a list of the transformation 
properties of the coefficients under discrete symmetries.
The T and CPT transformation properties hold up to non-hermitian 
contributions to the decay amplitude. This table tells us that the 
CP-invariance relation $b_{+}=b_{-}$ can only be violated if CP is not 
conserved  (i.e., $a_{f}\tilde{a}_{f}\neq 0$) and if absorptive parts of the 
amplitude are taken into account. In fact, the functions $b_{\pm}$ are zero 
to leading order for model (\ref{Ly}). Since the Born decay matrix is 
hermitian, we get from CPT invariance the relation $b_{\pm}=b_{\mp}$. 
Further the interaction (\ref{Ly}) is C-conserving, which implies 
$b_{\pm}=-b_{\mp}$ and hence $b_{\pm}=0$. On the other hand the
CP-invariance relation $c_{3}=0$ can be -- according to Table 1 --
violated and is, in fact, violated by (\ref{Ly}) already at tree level.\par 
We will now discuss some general properties of the observables  (\ref{obs}).
Since the observable  ${\cal O}_{0}$ is C-odd, its expectation value is
zero at tree level and at arbitrary  order in the QCD or QED couplings. 
CP-violating absorptive parts generated by one and higher loop corrections 
induce a nonzero difference $b_{+}-b_{-}$ and therefore a nonzero 
expectation value of the observable ${\cal O}_{1}$. The CP-violating triple 
correlation $\<{\cal O}_{2}\>$ is generated by (\ref{Ly}) already at tree 
level  and receives at the next order a contribution from the CP-violating 
dispersive part of the one-loop decay amplitude. The other two observables 
 ${\cal O}_{3,4}$ measure CP-conserving spin-spin correlations.\par 
We will now  discuss the decay of a Higgs boson into top quark pairs, $f=t$
(Later we will also consider the case $f=\tau$.). We have calculated the 
density matrix $R$ to order $\alpha_s$. We use the following notation: 
\bea
R&=&R_{0}+\frac{\alpha_{s}}{\pi}R_{1}+O\!\Big(
\frac{\alpha_{s}^2}{\pi^2}\Big) \,\,,\nn \\
\Gamma&=&\Gamma_{0}+\frac{\alpha_{s}}{\pi}\Gamma_{1}+
O\!\Big(\frac{\alpha_{s}^2}{\pi^2}\Big) \,\,,\nn\\
\<{\cal O}_{i}\>&=&\<{\cal O}_{i}\>_{\samplen{0}}+\frac{\alpha_{s}}{\pi}
\<{\cal O}_{i}\>_{\samplen{1}}+O\!\Big(\frac{\alpha_{s}^2}{\pi^2}\Big)\,\, ,
\label{exp1}
\eea
where
\bea
\<{\cal O}_{i}\>_{\samplen{0}}&=&\frac{\Tr\{R_0\cdot{\cal O}_{i}\}}
{\Tr\{R_{0}\}} \,\,,\nn\\
\<{\cal O}_{i}\>_{\samplen{1}}&=&\frac{\Tr\{R_1\cdot{\cal O}_{i}\}}
{\Tr\{R_{0}\}}-\<{\cal O}_{i}\>_{\samplen{0}}\frac{\Gamma_{1}}{\Gamma_{0}}\,\,.
\label{exp2}
\eea
At Born level the rate and the expectation values are found to be:
\bea
\Gamma_{0}&=&\frac{N_{C}}{8\pi}\frac{m_{t}^2}{v^{2}}m_{\varphi}
\beta(a^{2}_{t}\beta^2+\tilde{a}^{2}_{t})\,\,,\nn\\[2ex]
\<{\cal O}_{1}\>_{\samplen{0}}&=&0\,\,, \nn\\
\<{\cal O}_{2}\>_{\samplen{0}}&=&\frac{-a_{t}\tilde{a}_{t}\beta}
{a^{2}_{t}\beta^{2}+\tilde{a}^{2}_{t}} \,\,,\nn\\
\<{\cal O}_{3}\>_{\samplen{0}}&=&\frac{a^{2}_{t}\beta^2-
3\tilde{a}^{2}_{t}}{4(a^{2}_{t}\beta^{2}+\tilde{a}^{2}_{t})} \,\,,\nn\\
\<{\cal O}_{4}\>_{\samplen{0}}&=&-\frac{1}{4} \,\,,
\label{borncoeff}
\eea
where $N_{C}=3$. At tree level the $t\bar{t}$-system is in a pure state. 
This can also be explicitly checked using (\ref{borncoeff}). Radiative 
corrections to the coefficients will in general lead to mixed states. \par
The calculation of the QCD corrections to the results (\ref{borncoeff}) 
proceeds as follows: We calculate separately the virtual and real 
corrections to order $\alpha_s$.  All singularities are treated by 
dimensional regularization. Soft (IR) singularities appear both in the 
virtual and real corrections; they cancel explicitly in the sum of the two 
in accordance with the Kinoshita-Lee-Nauenberg theorem. We note that the 
necessary integrations over the hard gluon momentum can be done
analytically for all coefficients of the density matrix. In order to write 
the results for $\Gamma_1$ and $\Tr\{R_1\cdot{\cal O}_{i}\}$ entering 
(\ref{exp1}) and  (\ref{exp2}) in a compact form, we define 
 $\omega\equiv(1-\beta)/(1+\beta)$. Further $C_F=4/3$, and $\mathrm{Li}_2(x)$
denotes the dilogarithmic integral. In the results below the top quark mass is 
defined in the on-shell scheme. For the order $\alpha_s$ contribution 
to the decay rate we find:
\bea
 &&\Gamma_{1}=\frac{m^2_t}{v^2}\frac{N_C C_F}{4\pi}m_\varphi
\left\{ \left(a^{2}_{t}\beta^{2}+{\tilde a}^{2}_{t}\right)
\left[\left(-2\beta+\ln(\omega)\left(1+\beta^{2}\right)\right)
\left(\frac{\ln (1+\omega)}{2}+\ln(1-\omega)\right)
\right.\right. \nn\\
 &&\left.\left.
+\!\left(1\!+\!{\beta}^{2}\right)\!\left(\frac{\mathrm{Li}_2(\omega^{2})}{2}+
\mathrm{Li}_2(\omega)\right)
\!+\!\frac{\left(13\beta^{4}\!+\!48\beta^{3}\!-\!34\beta^{2}\!-\!3\right)
\ln(\omega)-6\beta\left(1-7\beta^{2}\right)}{32\beta^{2}}\right]\right.\nn\\
 &&\left.+3\left(1-\beta^{2}\right)\frac{\left(\beta^{4}+6\beta^{2}+1\right)
\ln(\omega)+2\beta \left(1+{\beta}^{2}\right)}{32\beta^{2}}\,{\tilde a}^2_{t}
\right\} \,\,.
\label{rate}
\eea 
The result (\ref{rate}) reduces to the standard model result for $a_{t}=1$, 
$\tilde{a}_{t}=0$ and agrees with the one given in \cite{Braaten}. In Fig.1a-c
we plot the decay rate $\varphi\to t\bar{t}$ for different values of $a_{t}$ and
 $\tilde{a}_{t}$ as a function of the Higgs mass for $m_{t}=175$ GeV. We 
include the 1-loop running of $\alpha_s$ with five active flavours and 
$\Lambda_{QCD}=200$ MeV. In all plots the dashed line represents the Born 
result and the full line is the result to order $\alpha_s$. Fig 1a and 
Fig 1c depict the well known cases of a scalar and pseudoscalar 
 \cite{Djou} Higgs boson, respectively. For illustrating the case of a 
Higgs boson with indefinite CP parity we choose $a_{t}=(2/3)^{1/2}$ and 
 $\tilde{a}_{t}=(1/3)^{1/2}$. In the two Higgs doublet model this corresponds 
to $v_{1}=v_{2}$ and maximal mixing of the three neutral Higgs bosons with 
definite CP parity (see, e.g.\cite{BSP}). For a given Higgs mass a nonzero 
pseudoscalar coupling enhances the rate. Note that in general the QCD 
corrections are important.\par 
We now turn to the other observables defined in (\ref{obs}).
The QCD corrections to their expectation values are determined by 
(cf (\ref{exp2})):
\bea
 &&\hspace{-7cm}\Tr\{R_{1}\cdot{\cal O}_{1}\}=
\frac{m^2_t}{v^2}N_{C}C_F2\pi m^2_\varphi(1-\beta^2) a_{t}\tilde{a}_{t}\,\, ,
\label{o1}
\eea

\bea
 &&\Tr\{R_{1}\cdot{\cal O}_{2}\}=
\frac{m^2_t}{v^2}N_{C}C_F
m^2_\varphi\,a_{t}{\tilde a}_{t}\,
\bigg\{2\left(1+\beta^{2}\right)
\left[\mathrm{Li}_2(\omega^{2})-4\mathrm{Li}_2(\omega)\right]+
8\beta\ln(1-\omega)\nn\\
&&\qquad\qquad+2\left[\ln(\omega)\left (1+\beta^{2}\right)-2\beta\right]
\ln(1+\omega)-2\beta\left(1+2\beta^2\right)-\ln^2(\omega)
\left(1-\beta\right)^2\nn\\
&&\qquad\qquad+\ln(\omega)\left(1-2\beta-3\beta^2\right)\bigg\}\,\, ,
\label{o2}
\eea

\bea
 &&\Tr\{R_{1}\cdot{\cal O}_{3}\}=\frac{m^2_t}{v^2}
\frac{N_C C_F}
{96}\frac{m^2_\varphi}{\beta^3}\bigg\{
\left({a}^{2}_{t}\beta^{2}-3{\tilde a}^{2}_{t}\right)\!
\left[48(1+\beta^4)\!\left(\mathrm{Li}_2(\omega^{2})+2\mathrm{Li}_2(\omega)
\right)\right.\nn\\
 &&\qquad\left.-96(1-\beta^{2})\mathrm{Li}_2(\omega^{2})+48\beta^{2}
\left(\ln(\omega)(1+\beta^{2})\!-2\beta\right)\!\Big(2\ln(1-\omega)
+\ln(1+\omega)\Big)\right.\nn\\
 &&\qquad\left.+8(1-\beta^{2})\bigg(3\ln(\omega)
\Big(\ln(\omega)\!-4\ln(1+\omega)\Big)\!-\pi^2\bigg)+6\left(23+15\beta^{2}
\right)\beta\right.\nn\\
 &&\qquad\left.+3\ln(\omega)\left(7(1-\beta^2)^2-24\beta^2+48
\beta^3+16\right)\right]\nn\\
 &&\qquad+{\tilde a}^{2}_{t}
\left[48(1-\beta^2)(3+\beta^2)\left(2\mathrm{Li}_2(\omega)-
\mathrm{Li}_2(\omega^{2})-2\ln(\omega)\ln(1+\omega)\right)\right.\nn\\
 &&\qquad\left.+6\beta\left(69-4\beta^{2}-9\beta^{4}\right)+8\left({
\pi }^{2}-3\ln^2(\omega)\right)\!\left(({1+\beta}^{2})^{2}-4\right)
\right.\nn\\
 &&\qquad\left.+3\ln (\omega)\left(69-9\beta
^{6}-43\beta^{2}+15\beta^{4}\right)\right]\bigg\}\,\, ,
\label{o3}
\eea

\bea
&&\Tr\{R_{1}\cdot{\cal O}_{4}\}=\frac{m^2_t}{v^2}\frac{N_C C_F}{96}
\frac{m^2_\varphi}{\beta}\bigg\{
\left[-24\beta^2\left(1+\beta^2\right)\!\!
\left(\mathrm{Li}_2(\omega^2)+4\mathrm{Li}_2(\omega)\right)
-48\mathrm{Li}_2(\omega^2)\right.\nn\\
&&\qquad\left.+48\beta\left(1+\beta^2\right)\!\!\Big(\ln(\omega+1)
+2\ln(1-\omega)\Big)+12\ln^2(\omega)\left(1-\beta\right)^2\left(
1+\beta\right)^2\right.\nn\\
&&\qquad\left.-24\ln(\omega)\left(2(\beta^2+2\beta^4+1)\ln(1-\omega)
+\left(3+\beta^4\right)\ln(\omega+1)\right)-30\beta\left(1+\beta^2\right)
\right.\nn\\
&&\qquad\left.-4\pi^2\!\left(1-\beta^2\right)\!\!\left(1+2\beta^2\right)
+3\ln(\omega)\!\left(10\beta^2-\beta^4+15-24\beta(1+\beta^2)\right)\right]\!\!
\left(a^{2}_{t}\!+{\tilde a}^{2}_{t}\right)\nn\\
&&\qquad+24\left(1-\beta^2\right)\!\!\left[\left(1+\beta^2
\right)\!\!\left(\mathrm{Li}_2(\omega^2)+2\mathrm{Li}_2(\omega)\right)+
\Big(\ln(\omega+1)+2\ln(1-\omega)\Big)\right.\nn\\
&&\qquad\left.\times\!\left(\ln(\omega)(1+\beta^2)
-2\beta\right)-\ln(\omega)\!\left(1-\beta^2\right)
+3\beta\Big(1+\ln(\omega)\Big)
\right]\!\!\left(a^{2}_{t}\!-{\tilde a}^{2}_{t}\right)\bigg\}\,\, .
\label{o4}
\eea
The strikingly simple result obtained from (\ref{o1}) for $\<{\cal O}_1\>$ 
is plotted in Fig. 2, again for  $a_{t}=(2/3)^{1/2}$ and 
 $\tilde{a}_{t}=(1/3)^{1/2}$. Remember that a nonzero value for this 
 correlation signals CP violation induced by absorptive parts. Close to 
the $t\bar{t}$ threshold the effect is of the order of 20\%. Note that close 
to threshold the infrared singularities due to Coulomb gluons should be 
resummed and hence our result will be substantially modified by higher order 
corrections which we do not consider here. For Higgs masses around 400 GeV 
the correlation drops to about 10\%. To exhibit the dependence of the 
expectation values on the unknown model parameters $a_{t}$ and 
 $\tilde{a}_{t}$ we assume for the moment $a_{t}\tilde{a}_{t}\ge 0$ and define
\be
r_{t}=\frac{\tilde{a}_{t}}{a_{t}+\tilde{a}_{t}},
\label{r}
\ee 
which takes values $0\le r_{t}\le 1$. It is easy to see that the expectation 
values of the observables (\ref{obs}) depend only on $r_{t}$ and not 
separately on $a_{t}$ and $\tilde{a}_{t}$.
In Fig. 3 this dependence is shown for $\<{\cal O}_1\>$ and a Higgs 
mass $m_\varphi=400$ GeV. Around $r_{t}=0.3$ the value of $\<{\cal O}_1\>$ 
is about 11\%. The case $a_{t}\tilde{a}_{t}\le 0$ may be analysed in complete 
analogy by defining $\tilde{r}_{t}=\tilde{a}_{t}/(\tilde{a}_{t}-a_{t})$. 
Clearly, both cases can be distinguished by measuring the sign of  
 $\<{\cal O}_1\>$.\par
Fig. 4 shows the CP-violating triple correlation  $\<{\cal O}_2\>$ 
defined in (\ref{obs}) for the same set of parameters as used in Fig. 2. 
This correlation is nonzero at tree level and the QCD corrections are tiny. 
In contrast to  $\<{\cal O}_1\>$, it vanishes for $m_\varphi\!\to 2m_{t}$. 
For a given Higgs mass the correlation may reach values of $\pm 0.5$.  
The dependence of $\<{\cal O}_2\>$ on $r_{t}$ is shown in Fig. 5, again for 
 $m_{\varphi}=400$ GeV. We note that for all $r_{t}$ the order $\alpha_s$ 
corrections are again very small.\par 
We now discuss the results for the CP-even observables. For a scalar 
(pseudoscalar) Higgs particle, the expectation value $\<{\cal O}_3\>$ 
is $1/4\,(-3/4)$ at tree level. QCD corrections to these results are 
smaller than 1\% for all Higgs masses below 500 GeV and vanish for 
 $m_\varphi=2m_{t}$. The expectation value $\<{\cal O}_3\>$ is quite 
sensitive to the value of $r_{t}$, as shown in Fig. 6.\par
Finally, the  QCD corrections  to the expectation value $\<{\cal O}_4\>$, 
which is $-1/4$ for all values of $a_{t}$ and $\tilde{a}_{t}$, are at most 
2\% for $m_{\varphi}<500$ GeV. The dependence of these corrections on 
 $r_{t}$ is very weak as demonstrated in Fig. 7. This completes our 
discussion of the order $\alpha_s$ corrections to the $t\bar{t}$ density 
matrix (\ref{Rdef}). \par
For a Higgs boson with a mass smaller than $2 m_W$ the main decay modes are
 $\varphi\to b\bar{b},c\bar{c}$ and $\varphi\to \tau^-\tau^+$. Because of
hadronization effects, a spin analysis is difficult for $b$ and $c$ 
quark pairs -- but it is feasible for $\varphi\to\tau^-\tau^+$. We illustrate 
this below for the case of a Higgs boson with $m_{\varphi}=100$ GeV decaying 
into a pair of $\tau$ leptons. The QED result (again in the on-shell scheme) 
follows from the formulae (\ref{exp1})--(\ref{r}) by replacing 
\be
N_C\to 1, \quad C_F\to 1, \quad \alpha_s \to \alpha, \quad
m_{t}\to m_{\tau}, \quad
a_{t}\to a_{\tau},\quad \tilde{a}_{t}\to \tilde{a}_{\tau},\quad 
r_{t}\to r_{\tau}\,\,.
\ee
The partial decay rate $\Gamma(\varphi\to\tau^-\tau^+)$ to order $\alpha$ 
is $0.202$ MeV, which is 2.4\% smaller than the Born rate. This result is 
practically the same for all the three choices of the model parameters 
 $a_{\tau}$ and $\tilde{a}_{\tau}$ which were also used before in the case
 of $\varphi\to t\bar{t}$. 
Because of the factors $m^2_\tau/m^2_\varphi\sim O(10^{-3})$ and 
 $\alpha(m_{\varphi})\approx 1/128$, the correlation $\<{\cal O}_{1}\>$ is tiny and 
is therefore not a useful tool for analyzing the CP nature of a light 
Higgs boson. In contrast, the triple correlation $\<{\cal O}_{2}\>$ takes 
the large value $-0.47$ for $a_{\tau}=(2/3)^{1/2}$ and 
 $\tilde{a}_{\tau}=(1/3)^{1/2}$. Its dependence on $r_{\tau}$ is shown in 
Fig. 8. The QED corrections to this result are below 1\%. As in the QCD 
case, the CP-even $\tau^-\tau^+$ spin-spin correlation $\<{\cal O}_{3}\>$ 
is also sensitive to the value of $r_{\tau}$. This is shown in Fig. 9. 
Finally, the QED corrections to the value $\<{\cal O}_{4}\>=-1/4$ are 
below $0.3$\%. 

\section{CP asymmetries in angular correlations of the final states}

\setcounter{equation}{0}
In this section we investigate the prospects to determine the CP parity of
the Higgs boson decaying into  $t\bar{t}$ or  $\tau^-\tau^+$ pairs 
by measuring suitable angular correlations between the final state 
particles into which the fermions decay. As we have shown in the last 
section, the radiative corrections to normalized observables are of the 
order of a few percent. From now on we will therefore neglect these 
corrections and perform a leading order analysis of the CP asymmetries. 
In this approximation the rest frame of the Higgs boson is identical to 
the $f\bar{f}$ c.m. system. In other words, the relation 
 ${\bf k}_f=-{\bf k}_{\bar{f}}$ holds and will be used in constructing the 
CP asymmetries below. It is straightforward to generalize the correlations 
 ${\cal E}_{1}, {\cal E}_{2}, {\cal E}_{3}$ defined below to data samples 
where $f$ and $\bar{f}$ are not antiparallel.\par
We first discuss the case of a Higgs boson heavy enough to decay into  
 $t\bar{t}$ pairs. We define a sample ${\cal A}$ containing events where 
the top quark decays semileptonically and the top antiquark decays 
hadronically,
\be
{\cal A}:\,\left\{\,\,
\begin{array}{l}
t\to \ell^++\nu_{\ell}+b \\ 
\bar{t}\to W^-+\bar{b}\to q+\bar{q}'+\bar{b}\,\,.
\end{array}\right.
\label{sampleA}
\ee 
The sample ${\bar{\cal A}}$ is defined by the charge conjugated decay 
channels of the $t\bar{t}$ pairs,
\be
{\bar{\cal A}}:\,\left\{\,\,
\begin{array}{l}
t\to  W^++b\to \bar{q}+q'+b \\ 
\bar{t}\to \ell^-+\bar{\nu}_{\ell}+\bar{b} \,\,.
\end{array}\right.
\label{sampleB}
\ee 
Each of these samples contains a fraction of $2/9$ of all $t\bar{t}$ pairs.
If we dismiss $\tau $ leptons as spin analyzers for the top quark, 
the remaining fraction is $4/27$. \par
The above decay modes are especially suited to study correlations that
result from top spin-momentum correlations. 
From the hadronic decays of the $t\,(\bar{t}$) the momentum of the 
 $t\,(\bar{t}$) may be reconstructed on an event by event basis and the 
acompanying lepton of the $\bar{t}\,(t)$ decay serves as an excellent 
spin analyzer. The knowledge of the top momentum allows a reconstruction 
of the top rest system. In our approximation we have 
 ${\bf k}_{\bar{t}}=-{\bf k}_t$, which means that also the $t$ and $\bar{t}$ 
rest system is known in the case of (\ref{sampleA}) and (\ref{sampleB}), 
respectively. We may therefore define the CP-odd correlation 
\be
{\cal E}_1=\<\hat{\bf k}
_t\cdot \hat{\bf p}_{\ell^+}^*\>_{\sample{A}} 
+\<{\hat {\bf k}}_t\cdot {\hat {\bf p}}_{\ell^-}^*
\>_{\sampleb{A}}\,\, ,
\label{E1}\ee
where ${\hat{\bf p}}_{\ell^+}^*$ is the flight direction of ${\ell^+}$ 
in the top quark rest system, and ${\hat{\bf p}}_{\ell^-}^*$ is the 
flight direction of ${\ell^-}$ in the top antiquark rest system.
(In the more general case one should replace 
 $\hat{\bf p}_{\ell^+}^*,\hat{\bf p}_{\ell^-}^*\to\hat{\bf p}_{\ell^+},
  \hat{\bf p}_{\ell^-}$ defined in the $\varphi$ rest frame.)
To measure ${\cal E}_1$, the corresponding expectation values are taken 
separately with respect to the two samples  ${\cal A}$ and ${\bar{\cal A}}$. 
The correlation (\ref{E1}) traces the CP-odd spin asymmetry 
 $\<{\cal O}_1\>$.\par   
We further define the following triple correlation:
\be
{\cal E}_{2}=\<{\hat {\bf k}}_{t}\cdot ({\hat {\bf p}}_{\ell^{+}}^*
\!\times 
{\hat {\bf p}}_{\bar b}^*)\>_{\sample{A}}-
\<{\hat {\bf k}}_{t}\cdot ({\hat {\bf p}}_{\ell^-}^*\times 
{\hat {\bf p}}_{b}^*)\>_{\sampleb{A}}\,\, ,
\label{E2}\ee
where ${\hat {\bf p}}_{\bar b}^*$ is the unit momentum  of the ${\bar b}$ 
in the $\bar{t}$ rest system and ${\hat {\bf p}}_{b}^*$ is measured in 
the $t$ rest system. The triple correlation (\ref{E2}) probes the CP 
violating spin-spin correlation $\<{\cal O}_2\>$.\par
The calculation of the correlations (\ref{E1}), (\ref{E2}) involves a trace
over the spin spaces of $t$ and $\bar{t}$ of the form 
 $\Tr\{R\,\rho_t\otimes\rho_{\bar{t}}\}$. The decay density matrix 
 $\rho_t$ for semileptonic $t$ decays, which is needed for expectation 
values taken with respect to sample ${\cal A}$, is given by
\be
\rho_{t}(t\to\ell^+\nu_{\ell}b)=\frac{6x_+(1-x_+)}{1-3\kappa^2+2\kappa^3}
\bigg[1\!{\rm l}+\hat{\bf p}_{\ell^+}^*\!\cdot{\bm \sigma}_{t}
\bigg]dx_+\frac{d\Omega_{\ell^+}}{4\pi}\,\, ,
\label{gammal}
\ee
where $\kappa=m_W^2/m_t^2$ and $x_+=2E_{\ell^+}^*/m_t\in [\kappa,1]$ 
is the scaled energy of the lepton. Further we need the decay density 
matrix for hadronic $\bar{t}$ decays, in which the $\bar{b}$ analyzes the 
top antiquark spin. It reads: 
\be
\rho_{\bar{t}}(\bar{t}\to W^{-} \bar{b}\to q+\bar{q}'+\bar{b} )
=\bigg[1\!{\rm l}+
\frac{1-2\kappa}{1+2\kappa}\,\hat{\bf p}_{\bar{b}}^*\cdot{\bm \sigma}_{\bar{t}}
\bigg]\frac{d\Omega_{\bar{b}}}{4\pi}\,\, .
\label{gammab}
\ee
The corresponding decay density matrices for sample ${\bar{\cal A}}$ can be
obtained by replacing $x_+\!\to x_-$, 
 $\hat{\bf p}_{\ell^+}^*\!\to-\hat{\bf p}_{\ell^-}^*$, 
 $d\Omega_{\ell^+}\!\to d\Omega_{\ell^-}$, 
 ${\bm \sigma}_{t}\!\to{\bm \sigma}_{\bar{t}}$ in (\ref{gammal}), and 
replacing 
 $\hat{\bf p}_{\bar{b}}^*\to-\hat{\bf p}_{b}^*$,
 ${\bm \sigma}_{\bar{t}}\to{\bm \sigma}_{t}$,
 $d\Omega_{\bar{b}} \to d\Omega_{b}$,
 ${\bm \sigma}_{\bar{t}}\to{\bm \sigma}_{t}$ in (\ref{gammab}).
Using these decay matrices, we get for the correlations ${\cal E}_{1,2}$:
\bea
{\cal E}_{1}&=&\frac{2}{3}\<{\cal O}_1\> \nn\,\, ,\\
{\cal E}_{2}&=&\frac{8}{9}\cdot\frac{1-2\kappa}{1+2\kappa}\<{\cal O}_2\>\,\, .
\eea
The factor $(1-2\kappa)/(1+2\kappa)\approx 0.41$ is the spin analyzer quality
of the $b(\bar{b})$. The statistical sensitivities of these correlations 
on the unknown couplings $a_{t}$ and $\tilde{a}_{t}$ are determined by the 
mean square fluctuations of the observables. We have 
\bea
 \<(\hat{\bf k}_{t}\cdot\hat{\bf p}_{\ell^+}^*)^2\>_{\sample{A}}
 &=&\<(\hat{\bf k}_{t}\cdot\hat{\bf p}_{\ell^-}^*)^2\>_{\sampleb{A}}
=\frac{1}{3}\,\, ,\nn\\
 \<(\hat{\bf k}_{t}\cdot(\hat{\bf p}_{\ell^+}^*\!
 \times\hat{\bf p}_{\bar{b}}^*))^2\>_{\sample{A}}&=&
\<(\hat{\bf k}_{t}\cdot(\hat{\bf p}_{\ell^-}^*\!
 \times\hat{\bf p}_{b}^*))^2\>_{\sampleb{A}}=\frac{2}{9}\,\,. 
\label{w1w2}
\eea
In order to decide whether or not the Higgs boson is a CP eigenstate, it is
sufficient to observe a nonzero value of  ${\cal E}_{1}$ or  ${\cal E}_{2}$,
which implies a nonzero $\<{\cal O}_{1}\>$ or $\<{\cal O}_{2}\>$, 
respectively.\par
The number $N_{t\bar{t}}^{(1,2)}$ of $\varphi\to t\bar{t}$ events that are 
needed to establish a nonzero correlation  $\<{\cal O}_{1,2}\>$ with three 
standard deviation (s.d.) significance are given by:
\bea 
N^{(1)}_{t\bar{t}}&=&
9\cdot\frac{27}{4}\cdot\frac{3}{2}\cdot
\frac{1}{\<{\cal O}_{1}\>^2}\,\, , \nn\\
N^{(2)}_{t\bar{t}}&=&
9\cdot\frac{27}{4}\cdot\frac{9}{16}\cdot
\left(\frac{1+2\kappa}{1-2\kappa}\right)^2\!\cdot
\frac{1}{\<{\cal O}_{2}\>^2} \,\, .
\label{sig12}
\eea
Here we took into account only electrons and muons in the samples ${\cal A}$
and ${\bar{\cal A}}$. In Fig. 10 we plot these numbers as a function of 
 $r_{t}$ defined in (\ref{r}) for $m_{\varphi}=400$ GeV (again assuming 
without loss of generality $a_{t}\tilde{a}_{t}\ge 0$). For a given number of 
events we can read off from the figure the interval for $r_{t}$ which leads 
to a nonzero correlation with $3$ s.d. significance. For example,
values $0.18 \lesssim r_{t} \lesssim 0.52$ would generate a nonzero 
 ${\cal E}_{2}$ with $3$ s.d. significance in a data sample of 
 $N_{t\bar{t}}=1500$ events. If no effect is seen in this sample, this 
interval for $r_{t}$ is excluded.\par
The above CP studies may be complemented by considering the correlation
\be
{\cal E}_{3}=\<\hat{\bf p}_{\ell^+}^*\cdot
\hat{\bf p}_{\bar{b}}^*\>_{\sample{A}} 
+\<{\hat {\bf p}}_{\ell^{-}}^{*}\cdot{\hat {\bf p}}_{b}^*
\>_{\sampleb{A}}\,\, ,
\label{E3}
\ee 
which is related to the CP-even spin-spin correlation $\<{\cal O}_3\>$ by
\be
{\cal E}_{3}= \frac{8}{9}\cdot\frac{1-2\kappa}{1+2\kappa}\cdot
\<{\cal O}_3\>\,\, .
\ee
From $\<(\hat{\bf p}_{\ell^+}^*\cdot\hat{\bf p}_{\bar{b}})^2\>_{\sample{A}}=
 \<(\hat{\bf p}_{\ell^-}^*\cdot\hat{\bf p}_{b})^2\>_{\sampleb{A}}=1/3$ 
we get
\be
N^{(3)}_{t\bar{t}}=
9\cdot\frac{27}{4}\cdot\frac{27}{32}\cdot
\left(\frac{1+2\kappa}{1-2\kappa}\right)^2
\cdot\frac{1}{\<{\cal O}_{3}\>^2}\,\, .
\label{sig3}
\ee
This number is also plotted (dotted line) as a function of $r_{t}$ in Fig. 10.
A simultaneous measurement of ${\cal E}_{2}$ and ${\cal E}_{3}$ with 
 $N_{t\bar{t}}=1500$ would have a $3$ s.d. sensitivity to values of 
 $r_{t}$ between $0.18$ and $1$. \par
Higgs boson production processes where the $\varphi$ rest system can
be reconstructed include $e^+e^-\to Z \varphi$ and $e^+e^-\to e^+ e^-\varphi$
at a future linear collider \cite{Zerwas} and, if realizable,
$\mu^+\mu^-\to\varphi$. For the $W^+ W^-$ fusion process
$e^+e^-\to\nu_e\bar{\nu}_e\varphi$ and for $\varphi$ production at the
LHC the above observables should be used with momenta obtained in the 
laboratory frame \cite{BBra}. This decreases the sensitivity of the 
observables as compared to above.\par
Essentially the same analysis as above can be carried out for the 
case $\varphi\to\tau^{-}\tau^{+}$. 
This mode is of interest for a quantum number analysis of Higgs bosons
with mass $m_{\varphi} < 2 m_W$. For a wide range of model parameters
one typically expects in this case the branching ratio of
the $\varphi\to \tau^-\tau^+$ mode to be about 8 percent. For analyzing
 the $\tau$ spin several decay modes  can be used.
We discuss here only the decays
 $\tau^{\mp}\to\pi^{\mp}\nu_{\tau}(\bar{\nu}_{\tau})$ and 
 $\tau^{\mp}\to\rho^{\mp}\nu_{\tau}(\bar{\nu}_{\tau})$. The decay density 
matrices are of the form 
\be
\rho_{\tau^{\mp}}(\tau^{\mp}\!\!\to B^{\mp}\nu_{\tau}(\bar{\nu}_{\tau}))=
\bigg[1\!{\rm l}\pm c_{\B}\hat{\bf p}^{*}_{\B^{\mp}}\!\cdot
{\bm \sigma}_{\tau^{\mp}}\bigg]\frac{d\Omega_{\B^{\mp}}}{4\pi}\quad\quad
(B=\pi,\rho)\,\, ,
\ee
with $c_{\pi}=1$, $c_{\rho}=0.456$ and $\hat{\bf p}^{*}_{\B^{\mp}}$ is 
the direction of flight of $B^{\mp}$ in the $\tau^{\mp}$ rest system. 
The mode $\tau\to a_{1}$, respectively $\tau\to 3\pi$ can also be taken into
account, see \cite{BNO}. We will assume in the following that the 
 $\tau^{\mp}$ flight directions can be reconstructed for the above decays. 
This should be possible with some effort, as was done in the CP studies in 
$\tau^-\tau^+$ production at LEP \cite{O2,A2,Wermes} (see also \cite{Kuhn}).
Since the CP-odd spin asymmetry $\<{\cal O}_{1}\>$ is tiny for 
 $\varphi\to\tau^{-}\tau^{+}$, we only discuss 
the CP-violating spin-spin correlation $\<{\cal O}_{2}\>$ and the CP-even 
quantity $\<{\cal O}_{3}\>$. Table 2 summarizes our results for the 
different decay modes of the $\tau^{-}\tau^{+}$ pairs and the 
corresponding correlations of the final state momenta which are sensitive 
to $\<{\cal O}_{2}\>$ and $\<{\cal O}_{3}\>$.\par
To constrain the unknown parameter $r_{\tau}$, we combine all these decay 
channels. The number of $\varphi\to\tau^-\tau^+$ events to get a 3 s.d. 
significance for a nonzero $\<{\cal O}_{2}\>$ is:
\bea
N_{\tau^{-}\tau^{+}}^{(2)}&=&9\cdot\frac{9}{8}\cdot\frac{1}{\<{\cal O}_{2}\>^2}
\cdot\frac{1}{0.013 + 0.056\, c_{\rho}^2 + 0.064\, c_{\rho}^{4} }\,\, . 
\label{Ntau3}
\eea
 $N_{\tau^{-}\tau^{+}}^{(3)}$ is obtained from (\ref{Ntau3}) by the 
replacements $9/8\to27/16$ and $\<{\cal O}_{2}\>\to\<{\cal O}_{3}\>$.
Both numbers are plotted in Fig. 11 as a function of $r_{\tau}$.
According to Fig. 11 about 2000 $\varphi\to\tau^-\tau^+$ events
would be needed  to establish 
 $r_{\tau} = {\tilde a}_{\tau}/(a_{\tau}+{\tilde a}_{\tau}) >$ 0.36 also at
the 3$\sigma$ level.

\section{Conclusions}

The analysis of the decays $\varphi\to f{\bar f}$ made above shows that 
the spin-spin correlations ${\cal O}_2$ and ${\cal O}_3$, respectively 
the correlations ${\cal E}_2$ and ${\cal E}_3$ are useful tools for 
determining the CP properties of neutral Higgs bosons. 
If there is CP violation in the Higgs sector it could be established with
these observables in a direct way. For a light Higgs boson with mass
$m_{\varphi} < 2 m_W$ the decay $\varphi\to \tau^-\tau^+$ offers a good
possibility to determine whether or not $\varphi$ is a CP eigenstate.
We found that with 2000 $\varphi\to \tau^-\tau^+$ events the combined
use of ${\cal E}_2$ and  ${\cal E}_3$ yields a sensitivity at the
3$\sigma$ level to a range of scalar and pseudoscalar Yukawa
couplings corresponding to the ratio
$r_{\tau}={\tilde a}_{\tau}/(a_{\tau}+{\tilde a}_{\tau}) >$ 0.36.
The sensititvity can be improved by using also other $\tau$ decay modes
than those considered above.\par
For the case of $\varphi \to t{\bar t}$ we showed that the
order $\alpha_s$ QCD corrections to the single spin asymmetry
and to the spin-spin correlations are small. Again, a sizeable
pseudoscalar component of $\varphi$ can be traced with
the combined use of the correlations ${\cal E}_2$
and ${\cal E}_3$. We  found that
 $r_t = {\tilde a}_t/(a_t+{\tilde a}_t) >$ 0.18 could be established
as a 3$\sigma$ effect with
about 1500 reconstructed $\varphi\to t{\bar t}$ events.\par
If Higgs boson(s) will be found these quantum number analyses should
eventually be feasible with appropriately tuned new high luminosity 
colliders.
\bigskip

\newpage

\begin{center}
{\renewcommand{\baselinestretch}{0.2}
\begin{tabular}[c]{|c|c|c|c|c|c|}
\hline
  & C & P & CP & T & CPT \\
  &    &      &      & (Im${\cal T}=0$) & (Im${\cal T}=0$) \\
\hline
 $A$ & $A$ & $A$ & $ A$ & $ A$ & $ A$ \\
 $b_{\pm}$ & $-b_{\mp}$& $-b_{\pm}$& $ b_{\mp}$ & $ b_{\pm}$ & $ b_{\mp}$ \\
 $c_{1}$ & $c_{1}$ & $c_{1}$ & $ c_{1}$ & $ c_{1}$ & $ c_{1}$ \\
 $c_{2}$ & $c_{2}$ & $c_{2}$ & $ c_{2}$ & $ c_{2}$ & $ c_{2}$ \\
 $c_{3}$ & $c_{3}$ & $-c_{3}$ & $-c_{3}$ & $-c_{3}$ & $ c_{3}$ \\
\hline
\end{tabular}}\\[2ex]
{\bf Table 1:}
\begin{minipage}[t]{6cm}
\baselineskip 14pt
Transformation properties of the structure functions under discrete
symmetry transformations.
\end{minipage}\\[3cm]
\end{center}

\begin{center}
{\renewcommand{\baselinestretch}{1}
\small
\begin{tabular}{|c|c|c|}
\hline
Decay mode & Fraction of &  \\
     of    & $\varphi\to\!\tau^{-}\tau^{+} $& Correlations \\
  $\tau^{-}\tau^{+}$ pair &   events & \\
\hline
$\ds {\cal B}:\,\left\{
\begin{array}{l}
\tau^-\to\pi^-\nu_{\tau}\\[1ex]
\tau^+\to\pi^+\bar{\nu}_{\tau}\\
\end{array}\right.
$ & 0.013 & 
\rule[1ex]{0cm}{6ex}\begin{tabular}{c}
$\ds \<\hat{\bf p}_{\tau^-}\!\cdot(\hat{\bf p}_{\pi^+}^*\!\times
\hat{\bf p}_{\pi^{-}}^*)\>_{\,\sample{B}}=\frac{4}{9}\<{\cal O}_{2}\>$ \\
\rule[-2.2ex]{0cm}{7ex}
$\ds \<\hat{\bf p}_{\pi^+}^*\!\cdot\hat{\bf p}_{\pi^{-}}^*\>_{\,\sample{B}}
=-\frac{4}{9}\<{\cal O}_{3}\>$
\end{tabular}\\
\hline
$\ds {\cal C}:\,\left\{
\begin{array}{l}
\tau^-\to\rho^-\nu_{\tau}\\[1ex]
\tau^+\to\rho^+\bar{\nu}_{\tau}\\
\end{array}\right.
$ & 0.064 & 
\rule[1ex]{0cm}{6ex}
\begin{tabular}{c}
$\ds \<\hat{\bf p}_{\tau^-}\!\cdot(\hat{\bf p}_{\rho^+}^*\!\times
\hat{\bf p}_{\rho^{-}}^*)\>_{\,\sample{C}}=\frac{4}{9}
c_{\rho}^2\<{\cal O}_{2}\>$ \\
\rule[-2.2ex]{0cm}{7ex}
$\ds \<\hat{\bf p}_{\rho^+}^*\!\cdot\hat{\bf p}_{\rho^{-}}^*\>_{\,\sample{C}}
=-\frac{4}{9}c_{\rho}^2\<{\cal O}_{3}\>$
\end{tabular}\\
\hline
\begin{tabular}{l}
\rule{0cm}{5ex}$\ds {\cal D}:\,\left\{
\begin{array}{l}
\tau^-\to\rho^-\nu_{\tau}\\[1ex]
\tau^+\to\pi^+\bar{\nu}_{\tau}\\
\end{array}\right.$ \\[3ex]
\rule[-4ex]{0cm}{5ex}$\ds {\bar{\cal D}}:\,\left\{
\begin{array}{l}
\tau^-\to\pi^-\nu_{\tau}\\[1ex]
\tau^+\to\rho^+\bar{\nu}_{\tau}\\
\end{array}\right.$ 
\end{tabular} &
\begin{tabular}{c}
0.028\\[5ex]
0.028
\end{tabular} &
\rule[1ex]{0cm}{6ex}
\begin{tabular}{c}
$\ds \!\!\<\hat{\bf p}_{\tau^-}\!\cdot(\hat{\bf p}_{\pi^+}^*\!\times
\hat{\bf p}_{\rho^{-}}^*)\>_{\,\sample{D}}\!-
\<\hat{\bf p}_{\tau^-}\!\cdot(\hat{\bf p}_{\pi^-}^*\!\times
\hat{\bf p}_{\rho^{+}}^*)\>_{\,\sampleb{D}}=\frac{8}{9}c_{\rho}
\<{\cal O}_{2}\>$ \\
\rule[1ex]{0cm}{4ex}
$\ds \<\hat{\bf p}_{\pi^+}^*\!\cdot\hat{\bf p}_{\rho^{-}}^*\>_{\,\sample{D}}+
\<\hat{\bf p}_{\pi^-}^*\!\cdot\hat{\bf p}_{\rho^{+}}^*\>_{\,\sampleb{D}}
=-\frac{8}{9}c_{\rho}\<{\cal O}_{3}\>$
\end{tabular} \\
\hline
\end{tabular}}\\[2ex]
{\bf Table 2:}
\begin{minipage}[t]{14cm}
\baselineskip14pt
Different decay modes of the $\tau^{-}\tau^{+}$ pairs and the 
corresponding correlations of the final state momenta which are sensitive 
to $\<{\cal O}_{2}\>$ and $\<{\cal O}_{3}\>$.
\end{minipage}
\end{center}

\newpage
\newpage

\begin{center}


\unitlength 1cm
\begin{picture}(10,12)
\put(0,0){\psfig{figure=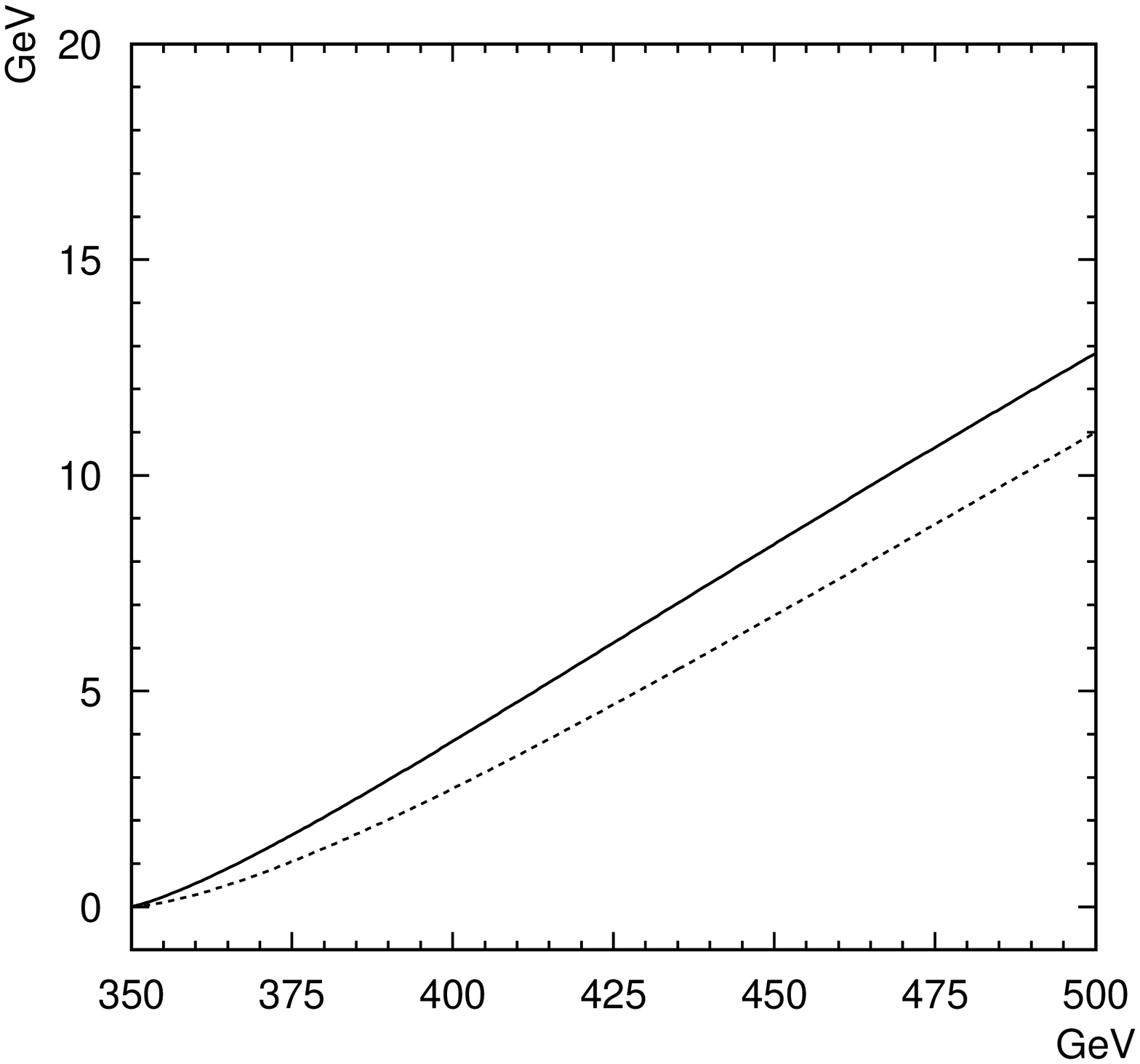,height=10cm,width=10cm}}
\put(0,-0.5){\small {\bf Figure 1a:}
\begin{minipage}[t]{10cm}
\baselineskip14pt
Decay rate $\varphi\to t\bar{t}$ in GeV as a function of the Higgs mass for 
 $m_t=175$ GeV, $a_{t}=1$ and $\tilde{a}_{t}=0$. The dashed line represents 
the Born result and the full line is the result to order $\alpha_s$.
\end{minipage}}
\end{picture}

\newpage

\unitlength 1cm
\begin{picture}(10,12)
\put(0,0){\psfig{figure=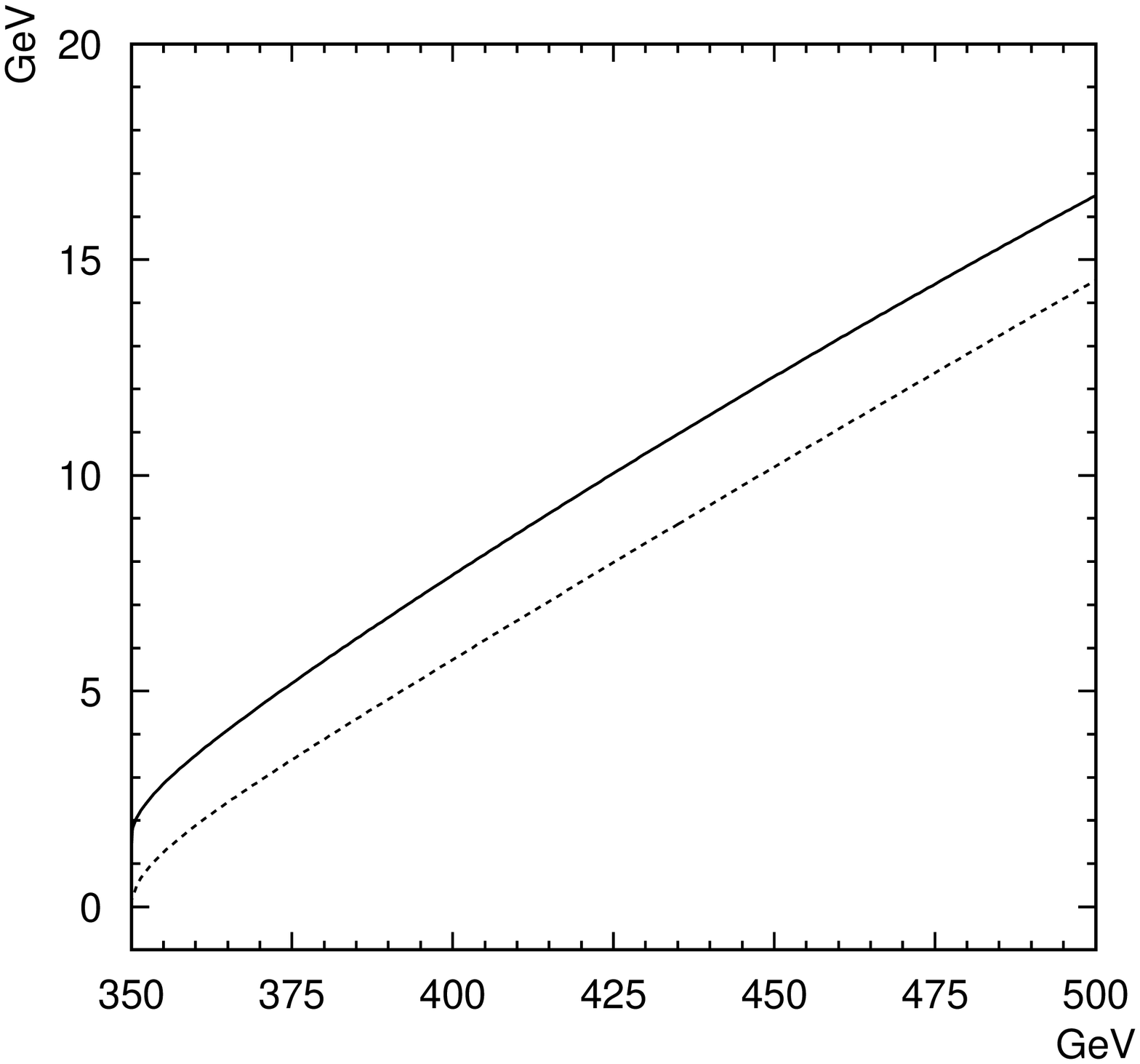,height=10cm,width=10cm}}
\put(0,-0.5){\small {\bf Figure 1b:}
\begin{minipage}[t]{10cm}
\baselineskip14pt
Decay rate $\varphi\to t\bar{t}$ in GeV as a function of the Higgs mass for 
 $m_t=175$ GeV, $a_{t}=(2/3)^{1/2}$ and $\tilde{a}_{t}=(1/3)^{1/2}$. The 
dashed line represents the Born result and the full line is the result to 
order $\alpha_s$.
\end{minipage}}
\end{picture}

\newpage

\unitlength 1cm
\begin{picture}(10,12)
\put(0,0){\psfig{figure=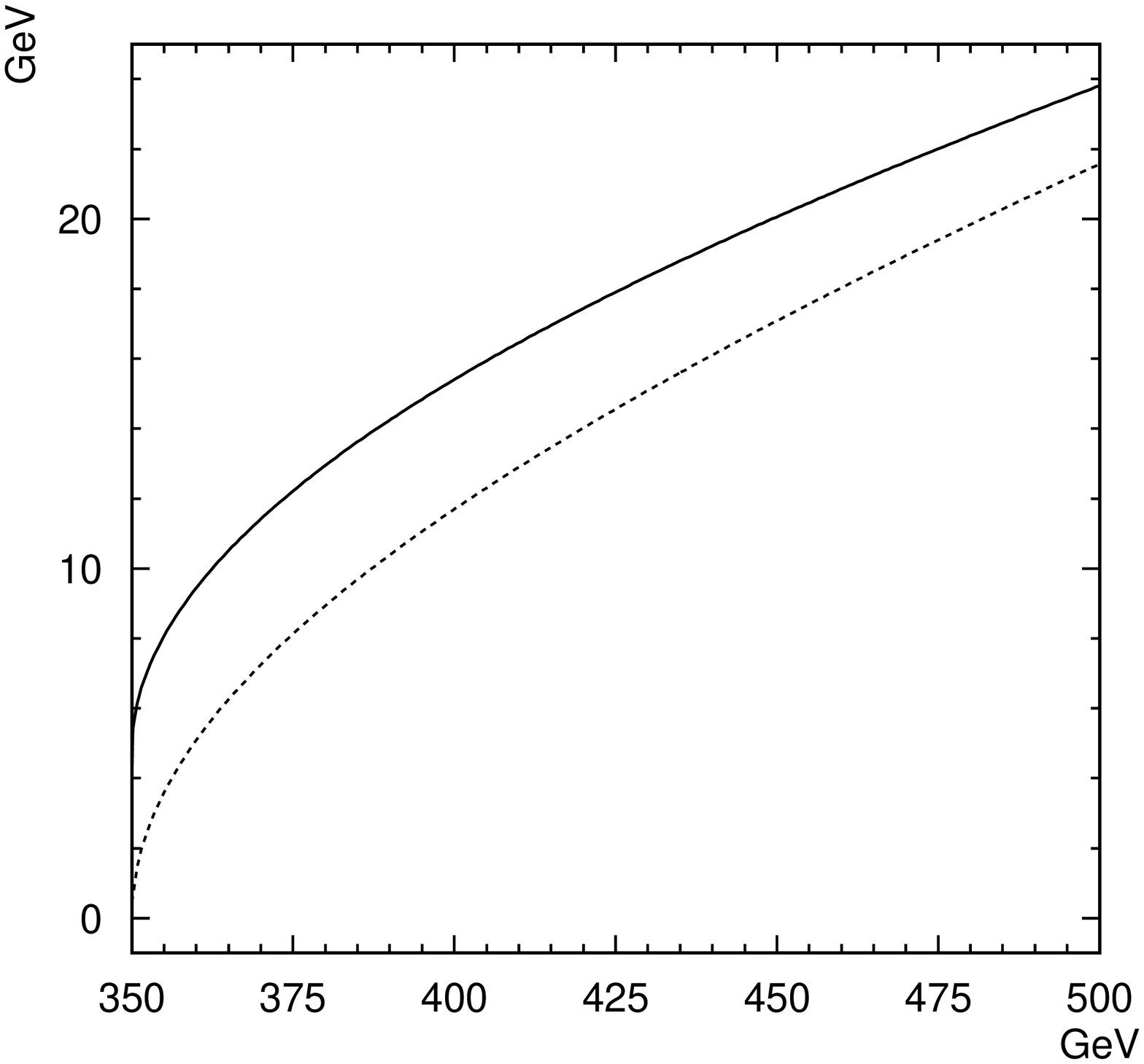,height=10cm,width=10cm}}
\put(0,-0.5){\small {\bf Figure 1c:}
\begin{minipage}[t]{10cm}
\baselineskip14pt
Decay rate $\varphi\to t\bar{t}$ in GeV as a function of the Higgs mass for 
 $m_{t}=175$ GeV, $a_{t}=0$ and $\tilde{a}_{t}=1$. The dashed line represents 
the Born result and the full line is the result to order $\alpha_s$.
\end{minipage}}
\end{picture}


\newpage

\unitlength 1cm
\begin{picture}(10,12)
\put(0,0){\psfig{figure=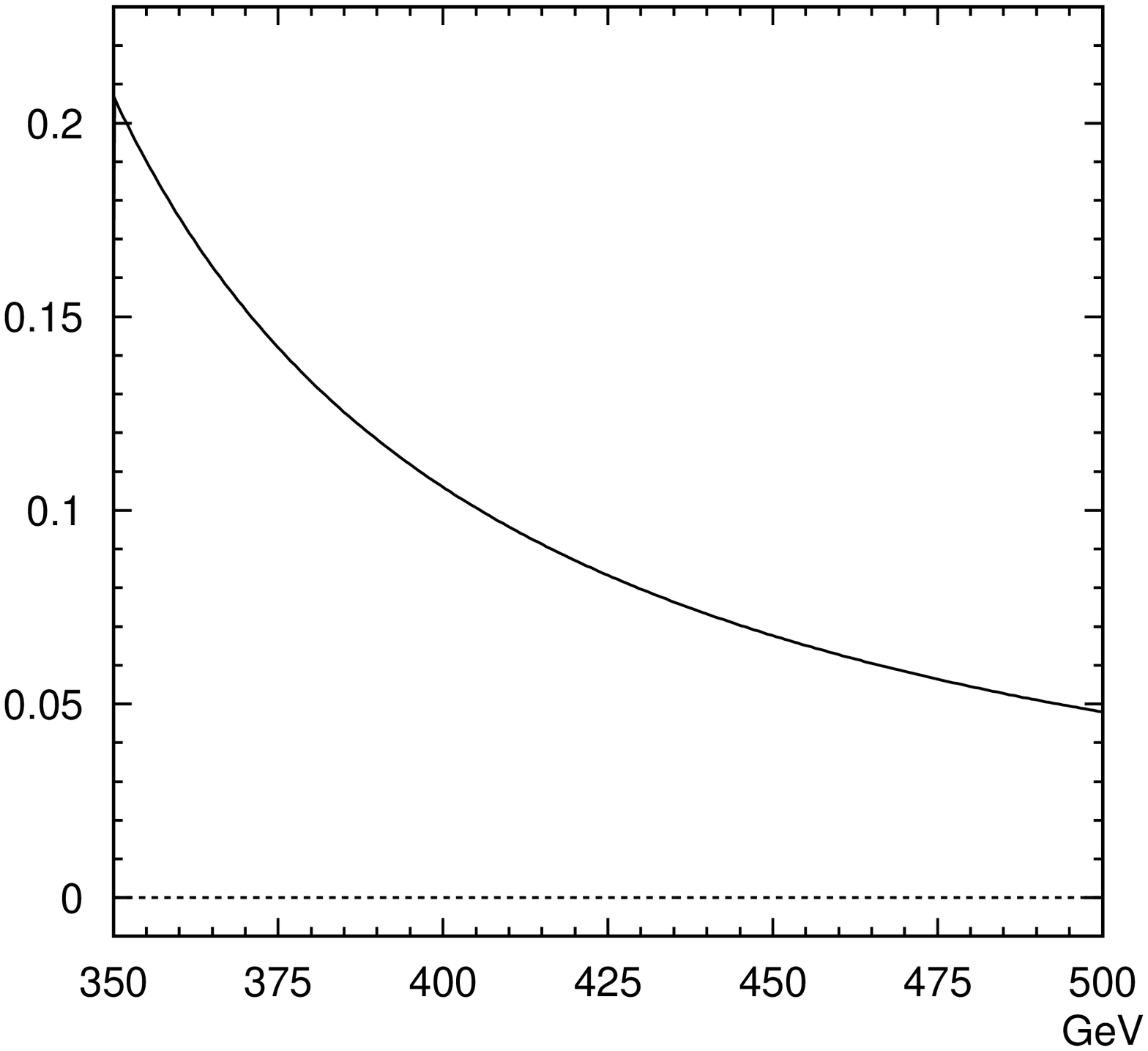,height=10cm,width=10cm}}
\put(0,-0.5){\small {\bf Figure 2:}
\begin{minipage}[t]{10cm}
\baselineskip14pt
Expectation value $\<{\cal O}_1\>$ as a function of the Higgs mass 
for $m_{t}=175$ GeV, $a_{t}=(2/3)^{1/2}$ and $\tilde{a}_{t}=(1/3)^{1/2}$. 
The dashed line represents the Born result and the full line is the result 
to order $\alpha_s$.
\end{minipage}}
\end{picture}

\newpage

\unitlength 1cm
\begin{picture}(10,12)
\put(0,0){\psfig{figure=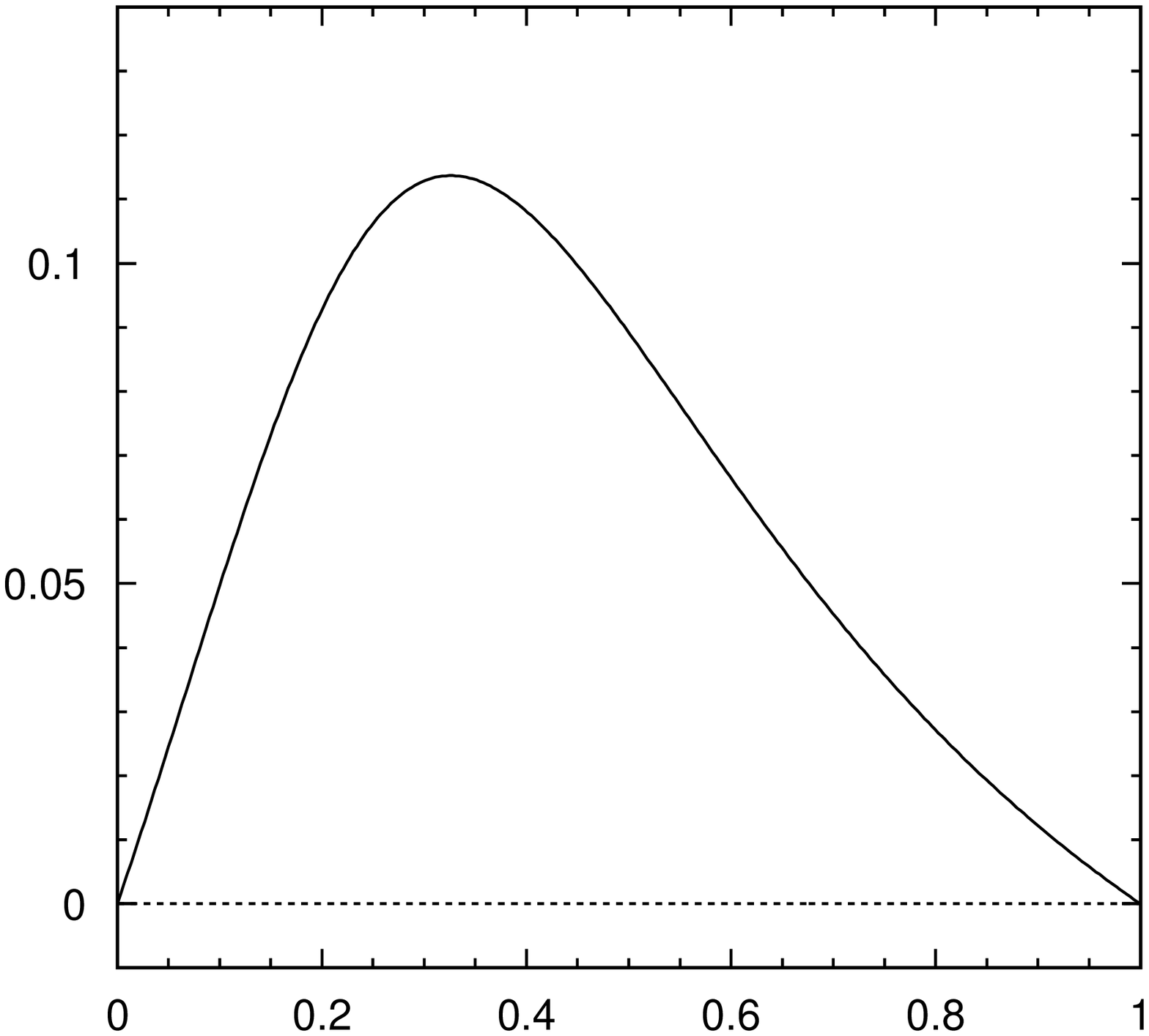,height=10cm,width=10cm}}
\put(0,-0.5){\small {\bf Figure 3:}
\begin{minipage}[t]{10cm}
\baselineskip14pt
Expectation value $\<{\cal O}_1\>$ as a function of 
$r_{t}=\tilde{a}_{t}/(a_{t}+\tilde{a}_{t})$ for fixed Higgs mass
 $m_\varphi=400$ GeV and $m_t=175$ GeV. The dashed line represents the 
Born result and the full line is the result to order $\alpha_s$.
\end{minipage}}
\end{picture}


\newpage

\unitlength 1cm
\begin{picture}(10,12)
\put(0,0){\psfig{figure=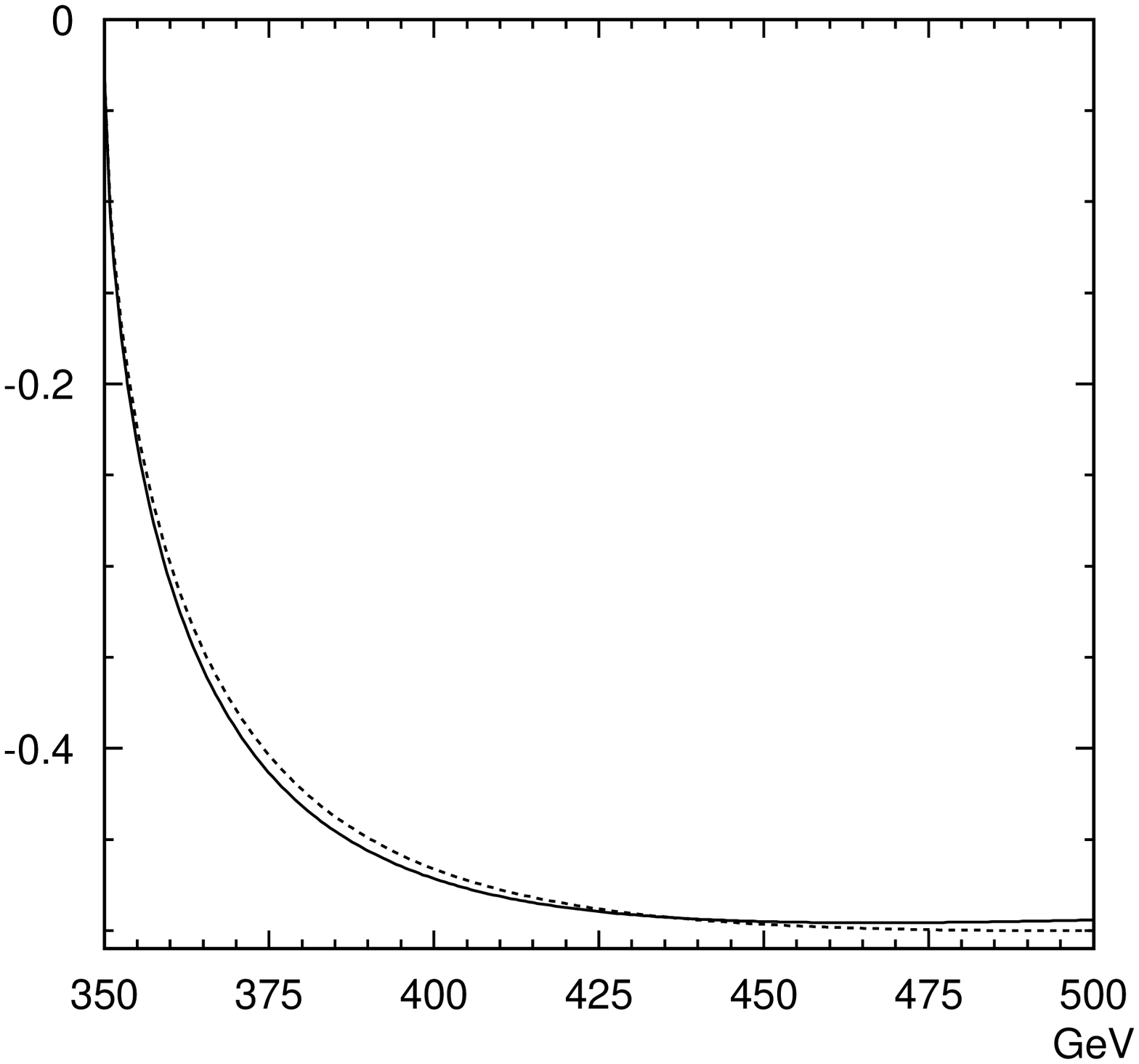,height=10cm,width=10cm}}
\put(0,-0.5){\small {\bf Figure 4:}
\begin{minipage}[t]{10cm}
\baselineskip14pt
Expectation value $\<{\cal O}_2\>$ as a function of the Higgs mass 
for $m_t=175$ GeV, $a_{t}=(2/3)^{1/2}$ and $\tilde{a}_{t}=(1/3)^{1/2}$. 
The dashed line represents the Born result and the full line is the result 
to order $\alpha_s$.
\end{minipage}}
\end{picture}

\newpage

\unitlength 1cm
\begin{picture}(10,12)
\put(0,0){\psfig{figure=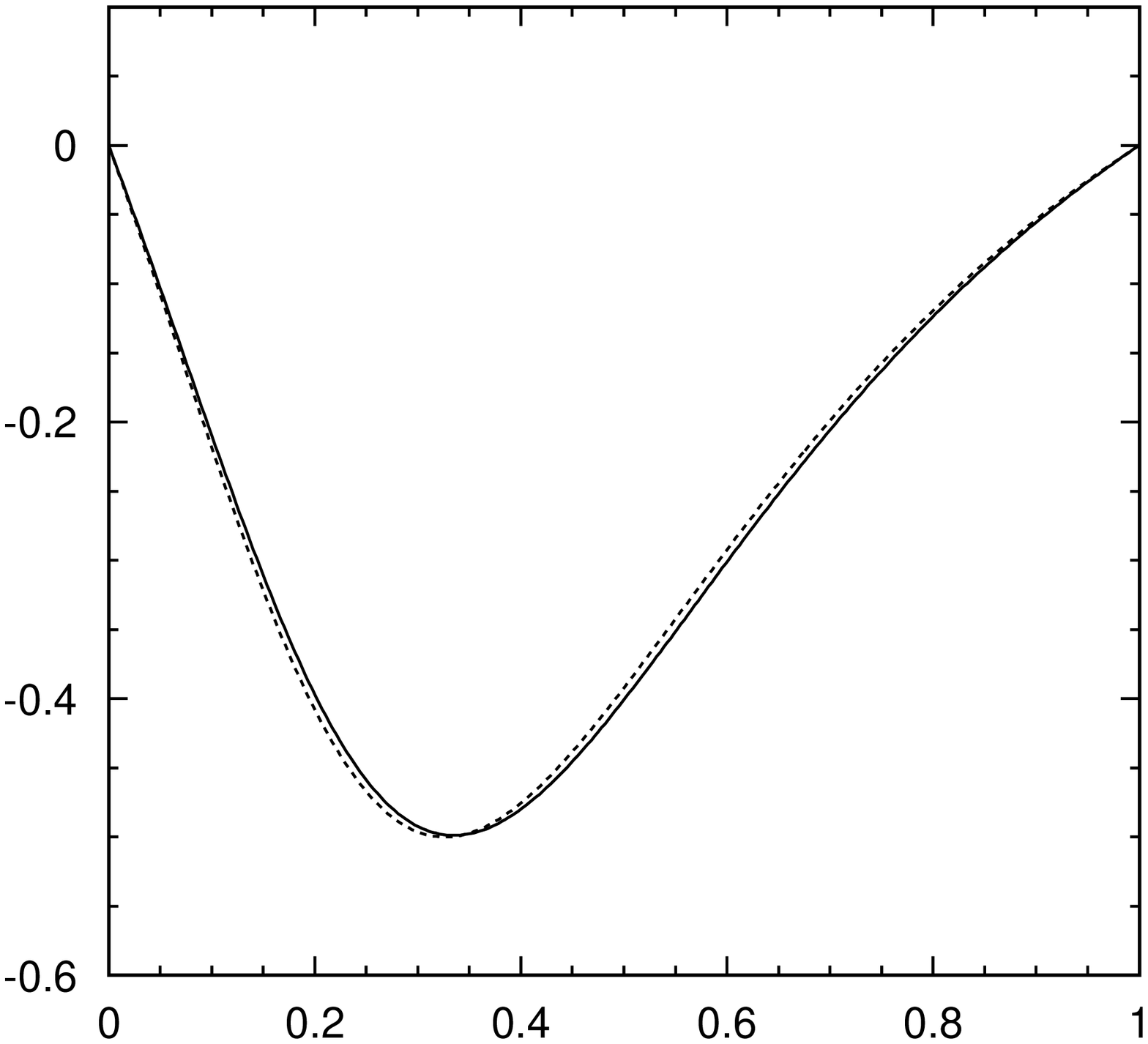,height=10cm,width=10cm}}
\put(0,-0.5){\small {\bf Figure 5:}
\begin{minipage}[t]{10cm}
\baselineskip14pt
Expectation value $\<{\cal O}_2\>$ as a function of 
$r_{t}=\tilde{a}_{t}/(a_{t}+\tilde{a}_{t})$ for fixed Higgs mass
 $m_\varphi=400$ GeV and $m_t=175$ GeV. The dashed line represents the 
Born result and the full line is the result to order $\alpha_s$.
\end{minipage}}
\end{picture}


\newpage

\unitlength 1cm
\begin{picture}(10,12)
\put(0,0){\psfig{figure=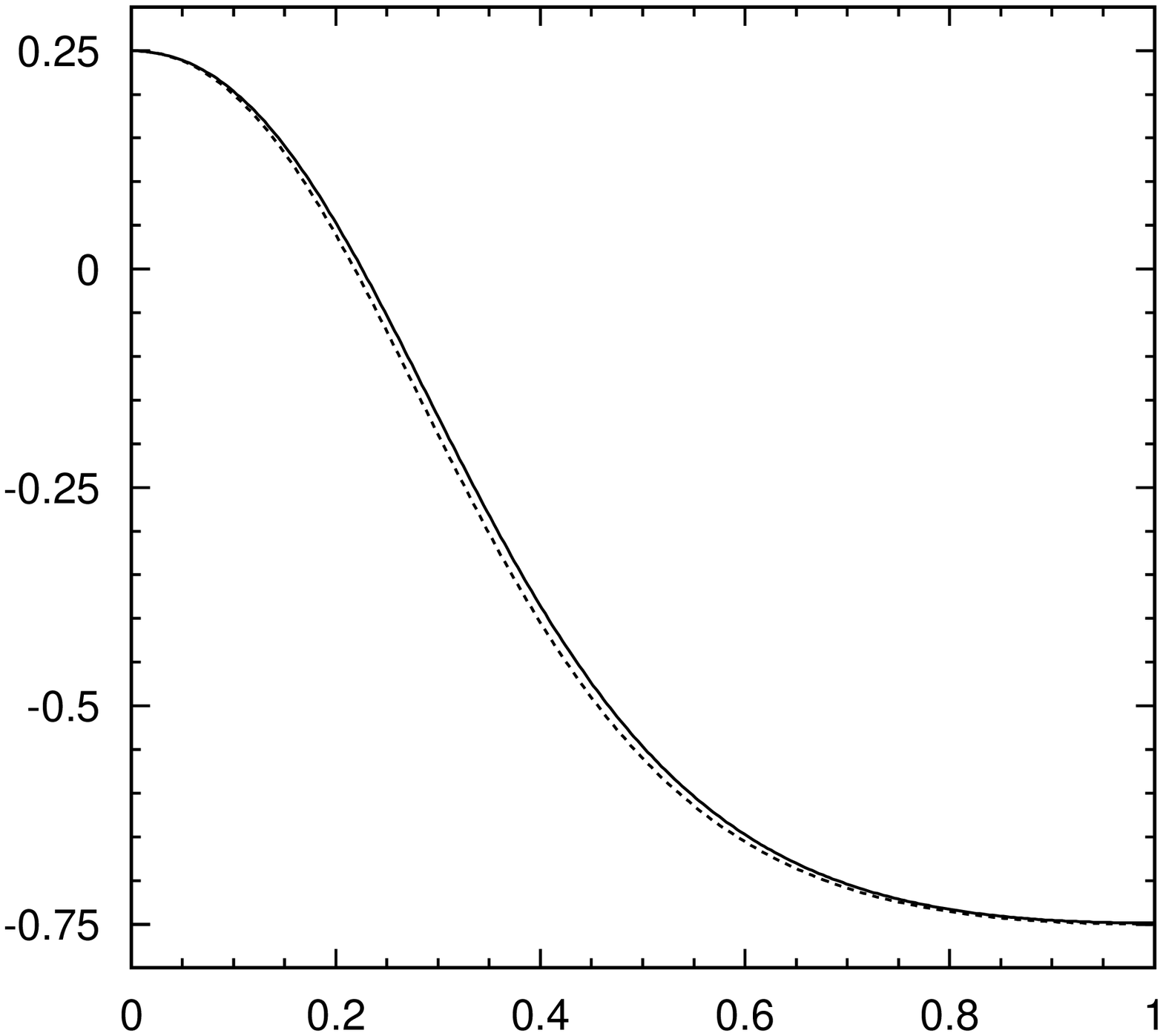,height=10cm,width=10cm}}
\put(0,-0.5){\small {\bf Figure 6:}
\begin{minipage}[t]{10cm}
\baselineskip14pt
Expectation value $\<{\cal O}_3\>$ as a function of 
$r_{t}=\tilde{a}_{t}/(a_{t}+\tilde{a}_{t})$ for fixed Higgs mass
 $m_\varphi=400$ GeV and $m_t=175$ GeV. The dashed line represents 
the Born result and the full line is the result to order $\alpha_s$.
\end{minipage}}
\end{picture}


\newpage

\unitlength 1cm
\begin{picture}(10,12)
\put(0,0){\psfig{figure=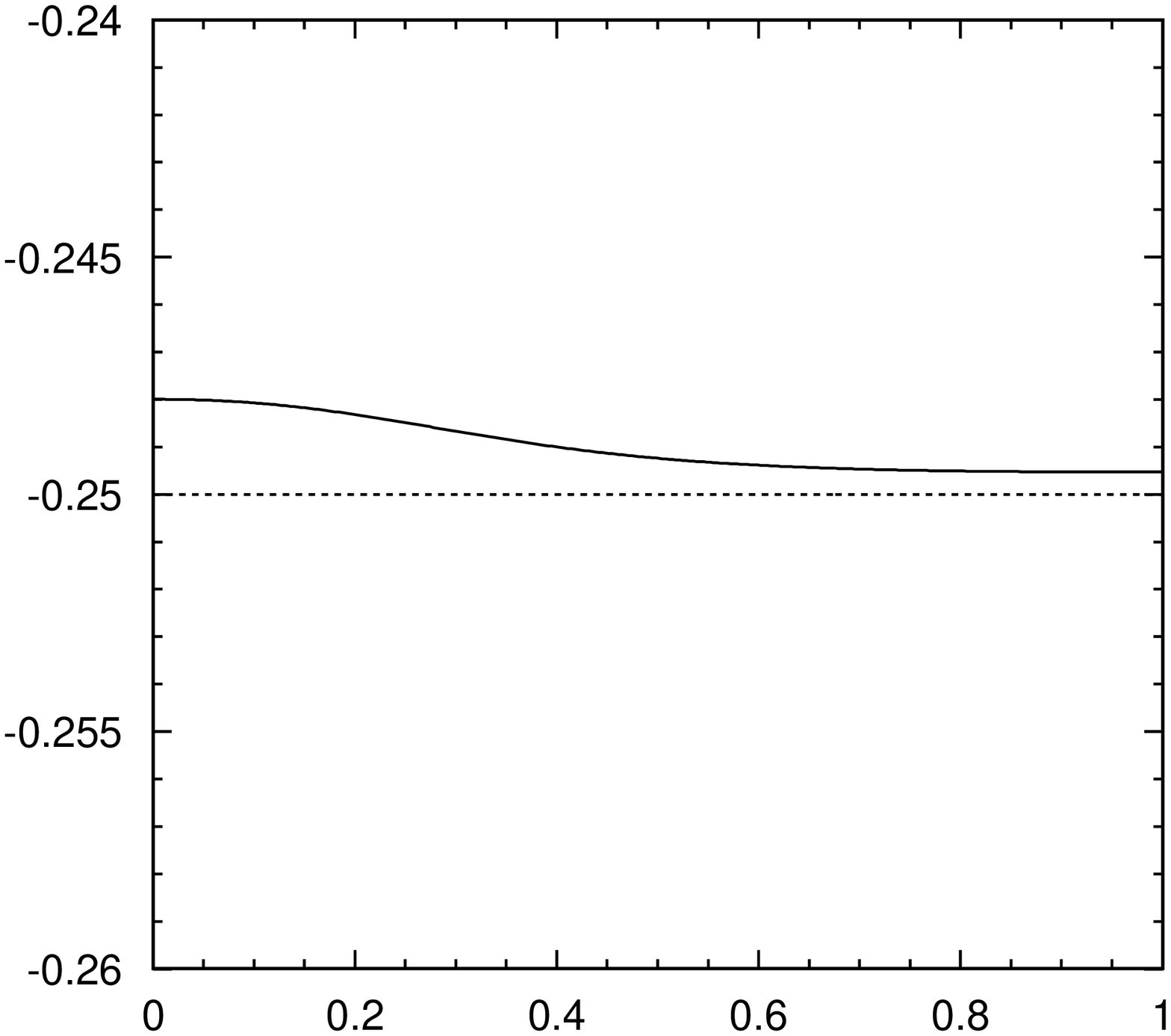,height=10cm,width=10cm}}
\put(0,-0.5){\small {\bf Figure 7:}
\begin{minipage}[t]{10cm}
\baselineskip14pt
Expectation value $\<{\cal O}_4\>$ as a function of 
 $r_{t}=\tilde{a}_{t}/(a_{t}+\tilde{a}_{t})$ for fixed Higgs mass
 $m_\varphi=400$ GeV and $m_t=175$ GeV. The dashed line represents 
the Born result and the full line is the result to order $\alpha_s$.
\end{minipage}}
\end{picture}


\newpage

\unitlength 1cm
\begin{picture}(10,12)
\put(0,0){\psfig{figure=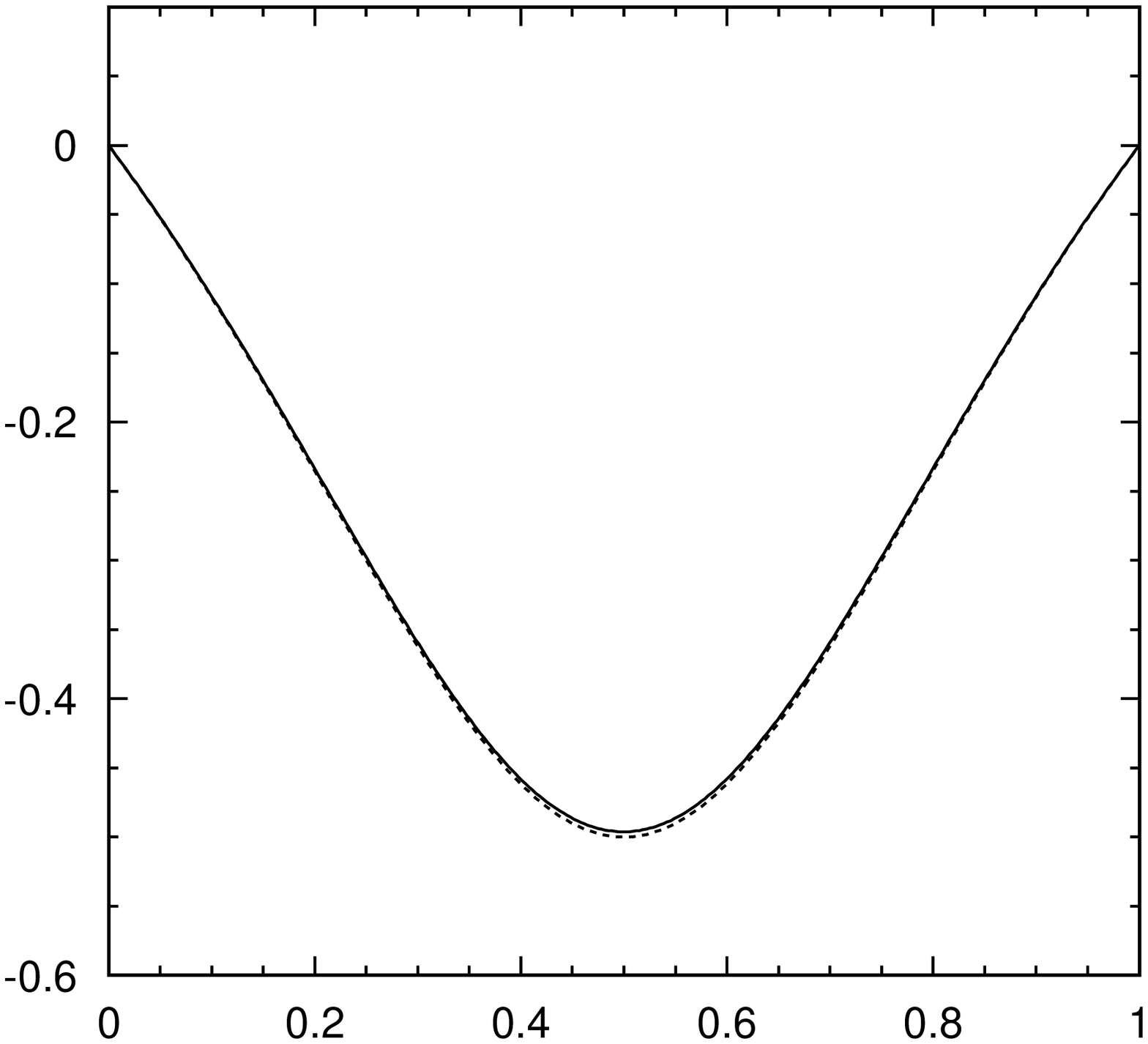,height=10cm,width=10cm}}
\put(0,-0.5){\small {\bf Figure 8:}
\begin{minipage}[t]{10cm}
\baselineskip14pt
Expectation value $\<{\cal O}_2\>$ for $\varphi\to\tau^-\tau^+$ as a function 
of $r_{\tau}=\tilde{a}_{\tau}/(a_{\tau}+\tilde{a}_{\tau})$ for fixed 
Higgs mass $m_\varphi=100$ GeV. The dashed line represents the Born result 
and the full line is the result to order $\alpha$.
\end{minipage}}
\end{picture}


\newpage

\unitlength 1cm
\begin{picture}(10,12)
\put(0,0){\psfig{figure=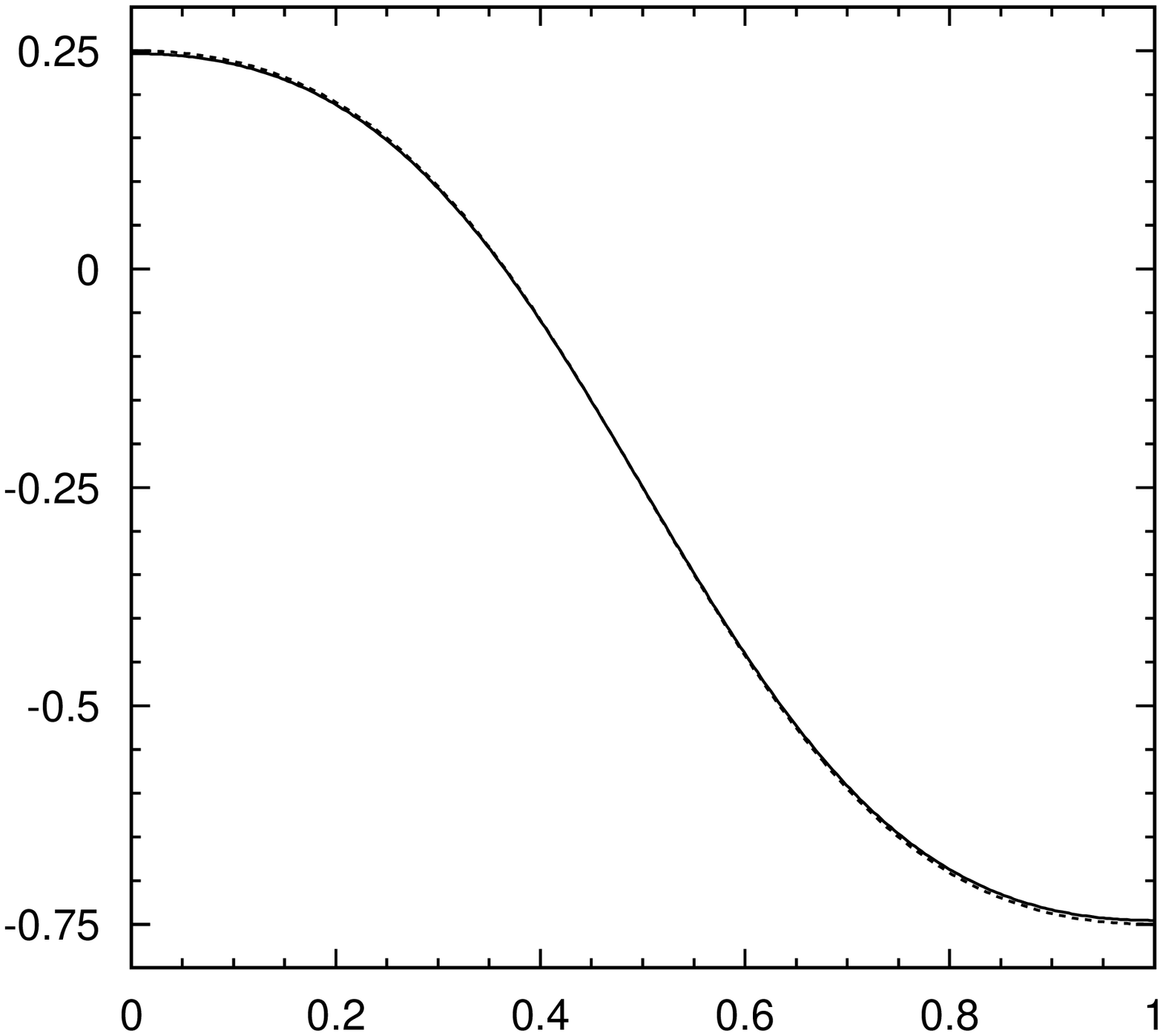,height=10cm,width=10cm}}
\put(0,-0.5){\small {\bf Figure 9:}
\begin{minipage}[t]{10cm}
\baselineskip14pt
Expectation value $\<{\cal O}_3\>$ for $\varphi\to\tau^-\tau^+$ as a function 
of $r_{\tau}=\tilde{a}_{\tau}/(a_{\tau}+\tilde{a}_{\tau})$ for fixed Higgs 
mass $m_\varphi=100$ GeV. The dashed line represents the Born result and 
the full line is the result to order $\alpha$.
\end{minipage}}
\end{picture}


\newpage

\unitlength 1cm
\begin{picture}(10,12)
\put(0,0){\psfig{figure=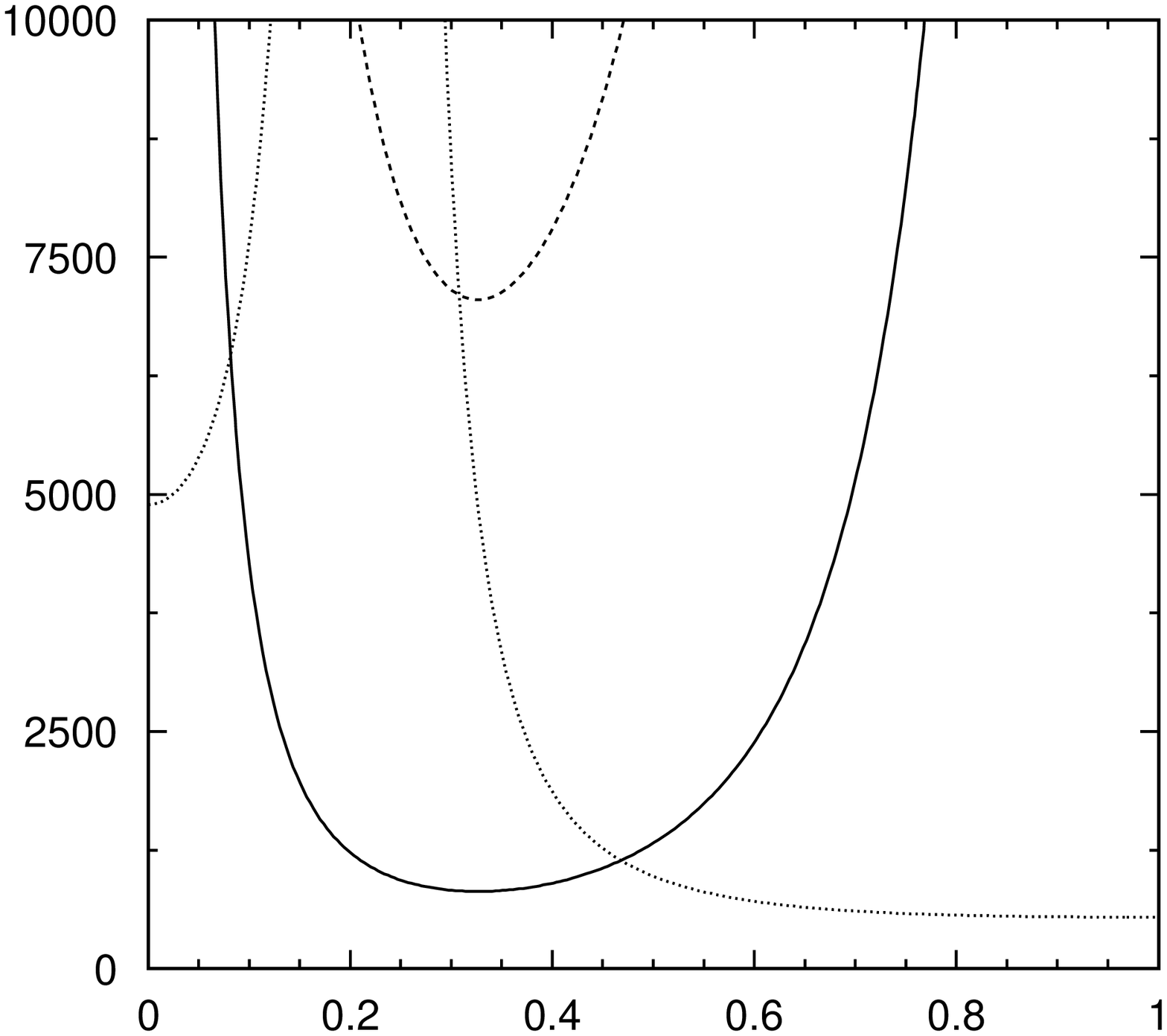,height=10cm,width=10cm}}
\put(0,-0.5){\small {\bf Figure 10:}
\begin{minipage}[t]{10cm}
\baselineskip14pt
Number of events $\varphi\to t\bar{t}$ to establish a nonzero 
correlation $\<{\cal O}_{1,2,3}\>$
(with 3 s.d. significance) as a 
function of $r_{t}=\tilde{a}_{t}/(a_{t}+\tilde{a}_{t})$ for fixed Higgs mass
 $m_\varphi=400$ GeV and $m_t=175$ GeV. The dashed line represents the result 
for $N^{(1)}_{t\bar{t}}$, the full line is the result for 
 $N^{(2)}_{t\bar{t}}$ and the dotted line is the result for 
 $N^{(3)}_{t\bar{t}}$.
\end{minipage}}
\end{picture}


\newpage

\unitlength 1cm
\begin{picture}(10,12)
\put(0,0){\psfig{figure=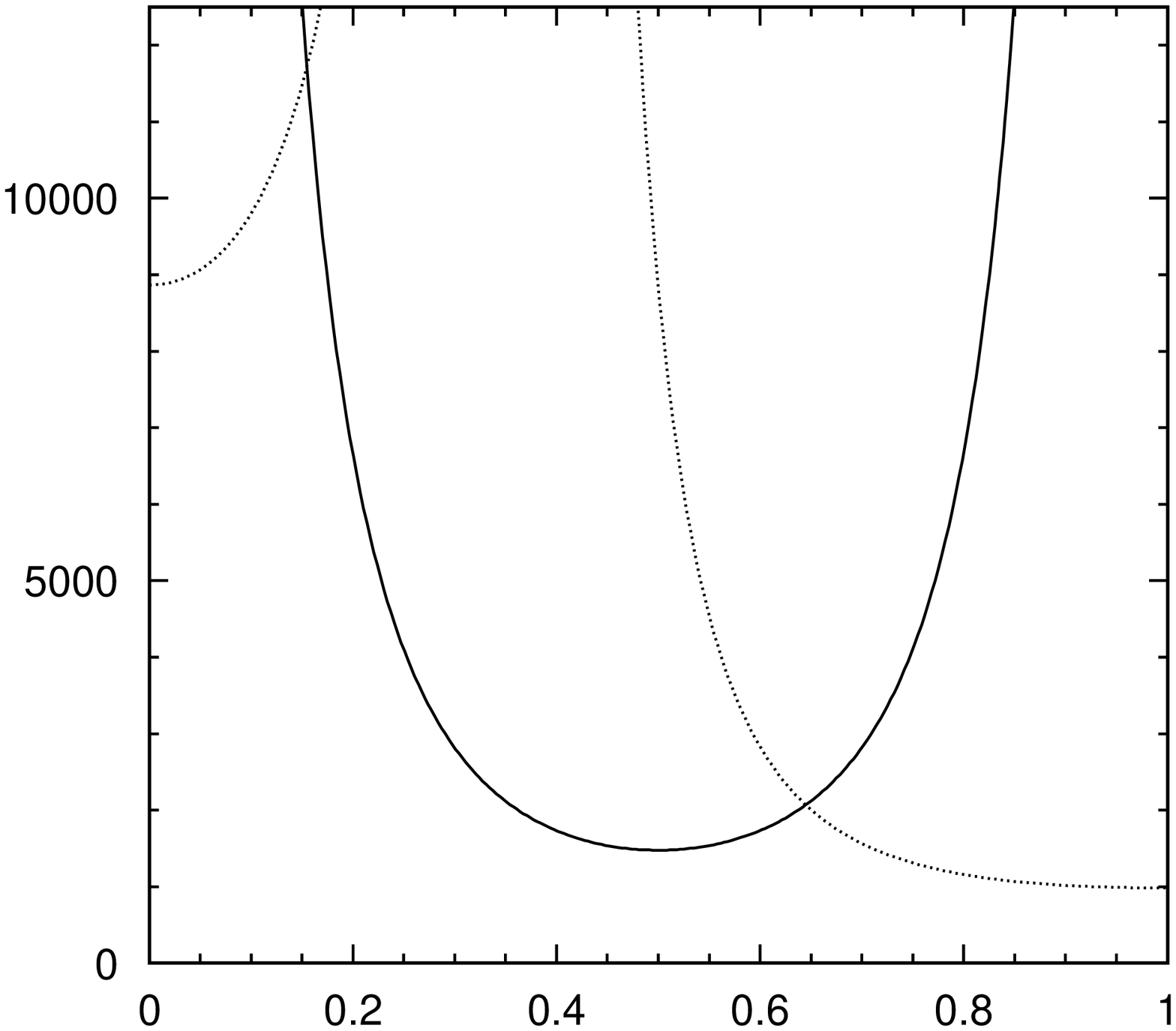,height=10cm,width=10cm}}
\put(0,-0.5){\small {\bf Figure 11:}
\begin{minipage}[t]{10cm}
\baselineskip14pt
Number of events $\varphi\to\tau^-\tau^+$ to establish a nonzero 
correlation $\<{\cal O}_{2,3}\>$
(with 3 s.d. significance) as a function of 
 $r_{\tau}=\tilde{a}_{\tau}/(a_{\tau}+\tilde{a}_{\tau})$ for fixed Higgs 
mass $m_\varphi=100$ GeV. The full line is the result for 
 $N^{(2)}_{\tau^-\tau^+}$ and the dotted line is the result for 
 $N^{(3)}_{\tau^-\tau^+}$.
\end{minipage}}
\end{picture}


\end{center}


\begin{thebibliography}{99}
\bibitem{comm}
In extensions of the SM such as 2 doublet extensions
CP violation in these modes is not a tree-level
but at best a one-loop effect and, hence, quite small.
\end{thebibliography}
\end{document}